\documentclass[sigplan,10pt,nonacm]{acmart}
\acmConference[Do not distribute]{Do not distribute}{December 03, 2021}{Stanford, CA, USA}
\acmYear{2021}
\acmISBN{} 
\acmDOI{} 
\startPage{1}

\setcopyright{none}

\usepackage[normalem]{ulem}
\usepackage{booktabs}   

\usepackage[nounderscore]{syntax}
\usepackage{xspace}
\usepackage{xparse}
\usepackage{microtype}
\usepackage{listings,multicol}
\usepackage{tikz}
\usepackage{colortbl}
\usepackage{float}
\usepackage{multicol}
\usepackage{amsbsy}
\usepackage{wrapfig}
\usepackage{mathtools}
\usepackage{enumerate}
\usepackage{enumitem}
\usepackage{varwidth}
\usepackage{multicol}
\usepackage{multirow}
\usepackage{tabularx}
\usepackage{outlines}
\usepackage[toc,page]{appendix}
\usepackage{amsmath}
\usepackage{pgfplots}
\usepackage{pgfplotstable}
\usepackage{xcolor,colortbl,xcolor-solarized}
\usepackage{cleveref}
\usepackage{tabu}
\usepackage{subcaption}
\usepackage{siunitx}

\usepackage{algorithm}
\usepackage[]{algpseudocode}

\colorlet{circgray}{black!65}
\newcommand{\circone}{{\color{circgray}\ding{202}}}
\newcommand{\circtwo}{{\color{circgray}\ding{203}}}
\newcommand{\circthree}{{\color{circgray}\ding{204}}}
\newcommand{\circfour}{{\color{circgray}\ding{205}}}
\newcommand{\circfive}{{\color{circgray}\ding{206}}}
\newcommand{\circsix}{{\color{circgray}\ding{207}}}
\newcommand{\circseven}{{\color{circgray}\ding{208}}}
\newcommand{\circeight}{{\color{circgray}\ding{209}}}

\newrobustcmd{\ubold}{\fontseries{b}\selectfont}
\pgfplotstableset{
    normpoint/.style={
        postproc cell content/.append code={
            \pgfkeysalso{@cell content/.add={\ubold}{}}
        },
    },
}

\tikzset{
        hatch distance/.store in=\hatchdistance,
        hatch distance=10pt,
        hatch thickness/.store in=\hatchthickness,
        hatch thickness=2pt
    }

    \makeatletter
    \pgfdeclarepatternformonly[\hatchdistance,\hatchthickness]{flexible hatch}
    {\pgfqpoint{0pt}{0pt}}
    {\pgfqpoint{\hatchdistance}{\hatchdistance}}
    {\pgfpoint{\hatchdistance-1pt}{\hatchdistance-1pt}}%
    {
        \pgfsetcolor{\tikz@pattern@color}
        \pgfsetlinewidth{\hatchthickness}
        \pgfpathmoveto{\pgfqpoint{0pt}{0pt}}
        \pgfpathlineto{\pgfqpoint{\hatchdistance}{\hatchdistance}}
        \pgfusepath{stroke}
    }
    \makeatother

\pgfplotscreateplotcyclelist{stallcycle}{%
      {fill=Paired-A},
      {fill=Paired-B},
      {fill=Paired-C},
      {fill=Paired-D},
      {fill=Paired-E},
      {fill=Paired-F},
      {fill=Paired-G},
      {fill=Paired-H}
    }

\usepackage{pifont}

\definecolor{todocolor}{rgb}{1.0,0.1,0.1}
\definecolor{keywordcolor}{rgb}{0.5,0,0.5}
\definecolor{textgray}{gray}{0.4}
\definecolor{mygray}{rgb}{0.5,0.5,0.5}
\lstdefinestyle{base}{
    xleftmargin=2em,
    columns=fullflexible,
    numbers=left,
    numbersep=5pt,
    keywordstyle=\color{solarized-blue},
    commentstyle=\color{solarized-base01},
    stringstyle=\color{solarized-magenta},
    numberstyle=\tiny\color{solarized-base0},
    emphstyle=[1]\color{solarized-red},
    emphstyle=[2]\color{solarized-green},
    emphstyle=[3]\color{solarized-magenta},
    basicstyle={\footnotesize\linespread{0.8}\color{solarized-base03}\ttfamily},
    escapeinside={(*}{*)}
  }
\lstdefinestyle{tight}{
    basicstyle={\scriptsize\linespread{0.8}\color{solarized-base03}\ttfamily},
  }
\lstdefinestyle{cppstyle}{
  style=base,
  language=C++,
  emph=[1]{environment,precompute,accelerate},
  emph=[2]{Tensor,IndexVar,IndexStmt},
  emph=[3]{Format},
}
\lstdefinestyle{spatial}{
    style=base,
    language=scala,
    emph=[1]{Foreach,Reduce,MemReduce,Scan,par,until,len,by,reduce},
    emph=[2]{FIFO,SRAM,load,store,stream_store_vec,enq,Reg,deq},
  }

\definecolor{todocolor}{rgb}{1.0,0.1,0.1}
\definecolor{keywordcolor}{rgb}{0.5,0,0.5}
\definecolor{textgray}{gray}{0.4}
\definecolor{mygray}{rgb}{0.5,0.5,0.5}

\lstdefinestyle{scalaStyle}{
  language=Scala,
  numberstyle=\scriptsize\color{nugray},
  basicstyle=\ttfamily\scriptsize,
}

\definecolor{myblue}  {RGB}{3,122,235}
\definecolor{mypurple}{RGB}{176,095,183}
\definecolor{myorange}{RGB}{252,128,8}
\definecolor{mygreen} {RGB}{0,143,0}
\definecolor{myred}   {RGB}{231,091,093}
\definecolor{mymaroon}   {RGB}{175,012,035}
\definecolor{mygray}  {RGB}{234,234,241}
\definecolor{nugray}  {RGB}{220,220,227}

\definecolor{mydarkgray}  {RGB}{80,80,80}

\newcommand{\HIDE}[1]{}

\newcommand{\pluseq}{\mathrel{+}=}

\newcommand\defeq{\mathrel{\overset{\makebox[0pt]{\mbox{\normalfont\scriptsize\sffamily def}}}{=}}}

\newcommand\code[1]{\lstinline[mathescape=true,basicstyle=\ttfamily\normalsize]|#1|}

\algnewcommand{\LineComment}[1]{\State // #1}

\newcommand{\bnfdef}{\mathrel{::=}}
\newcommand{\bnfalt}{\mathrel{\mid}}
\newcommand{\mT}{\mathcal{T}}

\newcommand{\iterFn}{lowerIter}

\colorlet{emititerationcolor}{myblue}
\colorlet{emitmappingcolor} {mypurple}
\colorlet{emitassemblycolor} {mygreen}
\colorlet{emitcomputecolor}  {mygreen}
\newcommand\emititerationcolor[1]{\textcolor{emititerationcolor}{#1}}

\colorlet{emititerationrefcolor}{emititerationcolor}
\colorlet{emitmappingrefcolor}  {emitmappingcolor}
\colorlet{emitcomputerefcolor}  {emitcomputecolor}
\colorlet{emitassemblyrefcolor} {emitassemblycolor}

\newcommand\EmitPseudo[2]{
  \expandafter\newcommand\csname #1\endcsname{%
    \textbf{emit} #2
  }
}

\EmitPseudo{initIters}{initialize iterators}
\EmitPseudo{initIterMeta}{initialize iterator metadata}
\EmitPseudo{loopHeader}{loop header}
\EmitPseudo{loopFooter}{loop footer}
\EmitPseudo{accessIters}{access iterators}
\EmitPseudo{resolveCoord}{resolve the coordinate of $i$}
\EmitPseudo{locateLocators}{locate from locators }
\EmitPseudo{condHeader}{conditional header}
\EmitPseudo{condFooter}{conditional footer}
\EmitPseudo{advanceIters}{advance iterators}

\EmitPseudo{mapTo}{map candidate coordinates to the original space}
\EmitPseudo{mapFrom}{map resolved coordinate to each derived space}

\newcommand{\ldotspack}{.\hskip-.5ex.\hskip-.5ex.}
\EmitPseudo{denseIter}{\texttt{Foreach or Reduce(\ldotspack{}=> i\ldotspack{})}}
\EmitPseudo{interIter}{\texttt{Foreach(Scan(\ldotspack{}or\ldotspack{}=> i\ldotspack{})}}
\EmitPseudo{unionIter}{\texttt{Foreach(Scan(\ldotspack{}and\ldotspack{}=> i\ldotspack{})}}
\EmitPseudo{sparseBVIter}{\texttt{Foreach(\ldotspack{}=> pos\ldotspack{})}}
\EmitPseudo{sparseIter}{\texttt{Foreach(Scan(\ldotspack{}=> i\ldotspack{})}}

\EmitPseudo{genBVtwo}{$\mathcal{B}_2=$ \Call{genBitvector}{$\mT_2$}}
\EmitPseudo{genBVone}{$\mathcal{B}_1=$ \Call{genBitvector}{$\mT_1$}}
\EmitPseudo{genBVResult}{scanner for result positions}

\EmitPseudo{computeCode}{compute code}
\EmitPseudo{assemblyCode}{assembly code}
\EmitPseudo{segmentInsert}{position insert code}

\algnewcommand\algorithmicswitch{\textbf{switch}}
\algnewcommand\algorithmiccase{\textbf{case}}
\algnewcommand\algorithmicdefault{\textbf{default}}

\algdef{SE}[SWITCH]{Switch}{EndSwitch}[1]{\algorithmicswitch\ #1\ \algorithmicdo}{\algorithmicend\ \algorithmicswitch}%
\algdef{SE}[CASE]{Case}{EndCase}[1]{\algorithmiccase\ #1}{\algorithmicend\ \algorithmiccase}%
\algdef{SE}[DEFAULT]{Default}{EndDefault}[1]{\algorithmicdefault\ }{\algorithmicend\ \algorithmicdefault}%
\algtext*{EndSwitch}%
\algtext*{EndCase}%
\algtext*{EndDefault}%

\algtext*{EndWhile}
\algtext*{EndIf}

\newcommand{\name}{Stardust\xspace}

\begin{document}

\title[Compiling Sparse Tensor Algebra to a Reconfigurable Dataflow Architecture]{\name: Compiling Sparse Tensor Algebra to a Reconfigurable Dataflow Architecture}         


\author{Olivia Hsu}
\affiliation{
  \institution{Stanford University}            
  \city{Stanford}
  \state{CA}
  \postcode{94305}
  \country{USA}                    
}
\email{owhsu@stanford.edu}          

\author{Alexander Rucker}
\affiliation{
  \institution{Stanford University}            
  \city{Stanford}
  \state{CA}
  \postcode{94305}
  \country{USA}                    
}
\email{acrucker@stanford.edu}          

\author{Tian Zhao}
\affiliation{
  \institution{Stanford University}            
  \city{Stanford}
  \state{CA}
  \postcode{94305}
  \country{USA}                    
}
\email{tianzhao@stanford.edu}          

\author{Kunle Olukotun}
\affiliation{
  \institution{Stanford University}            
  \city{Stanford}
  \state{CA}
  \postcode{94305}
  \country{USA}                    
}
\email{kunle@stanford.edu}          

\author{Fredrik Kj{\o}lstad}
\affiliation{
  \institution{Stanford University}            
  \city{Stanford}
  \state{CA}
  \postcode{94305}
  \country{USA}                    
}
\email{kjolstad@stanford.edu}          

\begin{abstract}
We introduce \name, a compiler that compiles sparse tensor algebra to reconfigurable dataflow architectures (RDAs).
\name introduces new user-provided data representation and scheduling language constructs for mapping to resource-constrained accelerated architectures.
\name uses the information provided by these constructs to determine on-chip memory placement and to lower to the Capstan RDA through a parallel-patterns rewrite system that targets the Spatial programming model. The \name compiler is implemented as a new compilation path inside the TACO open-source system.
Using cycle-accurate simulation, we demonstrate that \name can generate more Capstan tensor operations than its authors had implemented and that it results in 138$\times$ better performance than generated CPU kernels and 41$\times$ better performance than generated GPU kernels.
\end{abstract}



\maketitle

\section{Introduction}
\label{sec:intro}

Sparse tensor algebra (including linear algebra) is used in many applications, including data analytics~\cite{bader2008efficient}, computational science~\cite{asanovic2006landscape}, and machine learning~\cite{abadi2016tensorflow}. To accelerate sparse computation, hardware designers are developing specialized domain-specific accelerators~\cite{dadu2019towards,han2016eie,pal2018outerspace}.

Tensor algebra can be used to express a variety of computations. 
However, only a few functions, such as dense matrix multiplication and sparse matrix-vector multiplication, are used widely enough to warrant single-function hardware. 
This leaves a long tail of important tensor computations for which we cannot afford to build single-function hardware, but which would still benefit from performance and energy efficiency improvements beyond what is possible with CPUs and GPUs. Examples include tensor factorization in data analytics~\cite{bader2008efficient}, tensor contractions in computational chemistry~\cite{epifanovsky2013new}, and graph analytics expressed as linear algebra~\cite{mattson2013standards}. 
Kernel fusion enabled by compilers boosts arithmetic intensity~\cite{rucker2021capstan,zhao2019serving} and avoids unnecessary work~\cite{kjolstad2020sparse}.

To address the long tail of sparse functions that need to be accelerated, one promising approach is reconfigurable dataflow architectures (RDAs) with support for sparse operations~\cite{rucker2021capstan,hegde2019extensor}. RDAs provide general hardware for combining sparse data streams and for storing sparse temporary data structures on chip. They can be used to efficiently compute many different sparse linear and tensor algebra expressions. But to date, these architectures have only been programmed directly at the level of their sparse data combiners and physical memories. This means that programming them is not amenable to data scientists and machine learning software developers. Whereas to make RDAs much more accessible, we propose compiling to them directly from sparse tensor algebra expressions, along with a provided schedule, through a library API.

The challenge in compiling to sparse RDAs is managing their memories and controlling the combining of sparse streams of tensor coordinate-value pairs.
On CPUs and GPUs, the von Neumann architecture separates control logic from memories and presents the programmer with a convenient pull model---when you need data, you ask for it. 
But on RDAs the control logic is attached to memories in a push model of computation, where programmers explicitly manage data movement throughout the memory hierarchy. 
These challenges make RDAs harder to program than CPUs and GPUs, which is exacerbated by the complexity of sparse tensor algebra kernels.

Following Halide~\cite{halide2012} and TACO~\cite{senanayake2020}, we separate algorithms from schedules. And, like TACO, we further separate data representations~\cite{kjolstad2017tensor}. This design lets end users of tensor algebra focus on their application logic, using intuitive linear and tensor algebra expressions exposed as a library. These expressions can then be mapped to a CPU or, using the scheduling and data representation languages, to a dataflow architecture. Thus, end users and performance engineers can work separately. 
The experience from Halide~\cite{halide2012} has been that a clean scheduling language facilitates auto-scheduling research~\cite{ragankelley2013,mullapudi2016,adams2019,anderson2021} and lets performance experts tune code beyond the capabilities of current auto-scheduling systems.

In this paper, we describe how to compile arbitrary sparse tensor algebra expressions to Capstan~\cite{rucker2021capstan}, a recently published RDA. 
The key insight is to use separate languages to describe the tensor expression, the data representations and their placement on the RDA, and the mapping of the sparse iteration space of a tensor expression to sparse hardware.

\paragraph{Contributions}
\name is the first system that compiles sparse tensor algebra to a sparse RDA. Our contributions are:
\begin{itemize}
    \item a data representation language that can express accelerator tensor placement,
    \item a scheduling language that can express how portions of a (potentially transformed) sparse iteration space should be mapped to sparse accelerators,
    \item an algorithm that binds descriptions of data structures to specific physical memories, and
    \item a lowering rewrite system that maps sparse tensor algebra expressions to the declarative-sparse Spatial programming model for RDAs.
\end{itemize}
We implemented these concepts as a new compilation path in the open source TACO system~\cite{kjolstad2017tensor}, extending its data representation and scheduling languages to target dataflow hardware, as shown in \Cref{fig:overview}. We then use \name to compile a previously used benchmark set~\cite{kjolstad2017tensor} to the Capstan RDA~\cite{rucker2021capstan}. 
The data representation and scheduling languages in \name are sufficient to express optimized algorithms that outperform CPUs by 138$\times$ and GPUs by 41$\times$. 
Furthermore, the mapping languages enable new algorithms beyond those previously mapped to Capstan~\cite{rucker2021capstan}, demonstrating that the separation of algorithm, data representation, and scheduling provides a viable path for compilation to domain-specific architectures.

\section{Overview}
\label{sec:overview}

\begin{figure}
    \centering
    \includegraphics[width=\linewidth]{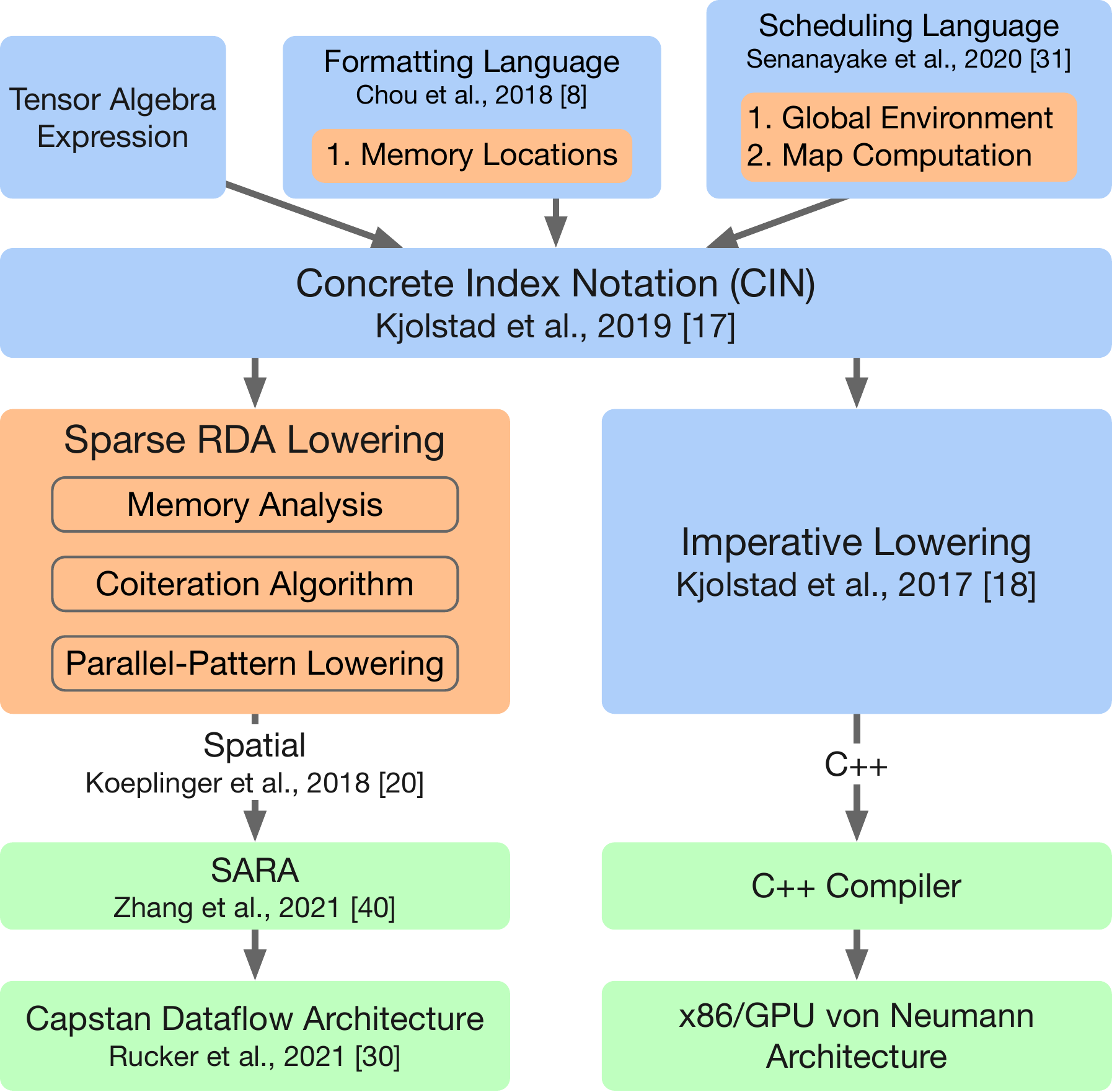}
    \caption{
        Overview of \name. Orange denotes new contributions, blue denotes prior work from the TACO literature, and green denotes the target compiler and hardware backends.
        \label{fig:overview}
        \vspace{1em}
    }
\end{figure}

\name takes three inputs: a tensor algebra expression, formats~\cite{chou2018}, and a schedule~\cite{senanayake2020}. 
These combine to produce an algorithm expressed in the \textit{concrete index notation} (CIN) intermediate representation (IR) of Kjolstad et al.~\cite{kjolstad2019}. 
\Cref{fig:overview} shows the relationship between our contributions (orange) and prior work on the TACO compiler (blue).
Unlike TACO, which generates C++~\cite{kjolstad2017tensor} and CUDA~\cite{senanayake2020} to target von Neumann machines, \name generates Spatial code~\cite{spatial} with parallel patterns to target a sparse RDA~\cite{rucker2021capstan}.

Domain-specific accelerators are by their nature more specialized than CPUs. And, although Capstan is more flexible than fixed-function hardware, it is still a specialized machine. To target specialized hardware, a user writes a schedule that reorganizes the computation until a sub-computation is exposed that maps directly onto the hardware.
Although TACO cannot map computation to specialized hardware, \name inherits its powerful sparse transformation framework.

\Cref{fig:overview} shows in orange the new format and scheduling language constructs we have added to \name that respectively map tensors and computation to the accelerator (\Cref{sec:format-schedule}). The memory location format construct lets the user denote that tensors live on the accelerator (\Cref{sec:coarse-mem}), while the memory analysis algorithm in the lowering component automatically determines the exact accelerator memory (\Cref{sec:mem-analysis}). And, the new \code{map} and \code{accelerate} commands let users map a sub-computation to the accelerator (\Cref{sec:schedule-map}), while the co-iteration algorithm automatically maps the pieces of the sub-computation to exact hardware implementations (\Cref{sec:coiteration-rewrite}).

\section{Background}
\label{sec:background}

Our work builds on two lines of prior work, as shown in \Cref{fig:overview}. We add format and scheduling constructs to the set of constructs proposed by Chou et al.~\cite{chou2018} and Senanayake et al.~\cite{senanayake2020} and extend the concrete index notation of Kjolstad et al.~\cite{kjolstad2019}. By compiling to Spatial, we can reuse the SARA~\cite{zhang2021sara} compiler to Capstan.

\subsection{Sparse Tensor Algebra Compilation}
\label{sec:taco-background}

The TACO compiler separates the algorithm (tensor index notation) from the tensor compression formats and computation transformations through the use of format~\cite{chou2018} and scheduling~\cite{senanayake2020} languages, respectively. It compiles sparse tensor algebra to imperative code by decomposing sparse iteration spaces into hierarchical set expressions of per-dimension data structures. Sparse algorithms are expressed in CIN (see \Cref{fig:concrete-index-notation-syntax}), which encodes iteration, computation, transformations, and temporary tensors~\cite{kjolstad2019}. Finally, TACO lowers CIN to generate efficient fused code that traverses irregular data structures by skipping unnecessary computation.

\paragraph*{Scheduling}
The sparse scheduling language proposed by Senanayake et al.~\cite{senanayake2020} provides a sparse iteration transformation framework. The framework modifies the sparse iteration space of an expression by taking its CIN statement and transforming it into a new CIN statement that represents a different algorithm for the same expression. The scheduling transformation framework describes optimizations to change the computation order, insert temporary tensors for partial sub-computation, exploit parallelism, and more. See \Cref{table:scheduling-old} for a reference to the TACO scheduling commands.

\begin{figure}
\footnotesize
\[
\hspace{-1em}
\begin{array}{rlrlrlrl}
\textit{Index Variable} & i & \textit{Index Variable List} & i* & \textit{Constants} & c & \textit{Tensors} & \mT\\
\end{array}
\]
\vspace*{-1.5em}
\[
\hspace{-1em}
\begin{array}{rlcl}
  \textit{Accesses} & a & \bnfdef & \mT(i*) \\
  \textit{Expressions} & e & \bnfdef & a \bnfalt c \bnfalt e + e \bnfalt e * e \bnfalt \ldots \\
  \textit{Statements} & S & \bnfdef & \forall_i~S \bnfalt a = e \bnfalt a \pluseq e \bnfalt \\ 
  & & & S~;~S \bnfalt S~\textsf{where}~S \bnfalt S~\textsf{s.t.}~r* 
  \\ 
  \textit{Scheduling Relation} & r & \bnfdef & \textsf{split\_up}(i, i_o, i_i, c) \bnfalt \textsf{split\_down}(i, i_o, i_i, c) \bnfalt \\
  & & & \textsf{fuse}(i_o, i_i, i_f) \bnfalt \ldots \\
\end{array}
\]
\caption{Concrete index notation (CIN) syntax.}
\label{fig:concrete-index-notation-syntax}
\end{figure}

\newcommand{\rasep}{{\hskip-1ex$\rightarrow$}}
 \begin{table}
  \centering
   \tabulinesep=2pt
  \footnotesize
   \begin{tabu} to 3.05in {@{}p{1.25in}@{}c@{\,\,}>{\scriptsize}X@{}}
    \toprule
    \rowfont{\footnotesize\sffamily\bfseries} Scheduling Commands & & Description\\
    \midrule
    \hangindent=2em\texttt{precompute}$(e, i*, i_w*, \mT)$         &\rasep & Inserts a \textsf{where} node to precompute a subexpression $e$ into a temporary tensor workspace $\mT$ with new indices $i_w*$ on the right-hand side of the newly introduced \textsf{where} node. \\
    \multicolumn{2}{c}{$\ldots \forall_{i*} a = e$\hfill$\xrightarrow[]{\textsf{precompute}(e, i*, i_w*, \mT)}$}  & $\ldots \forall_i a = \mT(i*)~\textsf{where}$ \\
    & & $\forall_{i_{w}*} \mT(i_w*)=e[i_{w}*/i*]$ \\
    \midrule
    \hangindent=2em\texttt{split\_up}$(i, i_o, i_i, c)$ &  \rasep& Stripmines a forall into a nested outer $i_o$ and constant-factor inner $i_i$ foralls.  \\
    \hangindent=2em\texttt{split\_down}$(i, i_o, i_i, c)$ &  \rasep& Stripmines a forall into a nested constant-factor outer $i_o$ and inner $i_i$ forall.  \\
    \multicolumn{2}{c}{$\ldots \forall_i S$\hfill$\xrightarrow[]{\textsf{split*}(i, i_o, i_i, c)}$}  & $\ldots \forall_{i_o} \forall_{i_i} S~\textsf{s.t. split*}(i, i_o, i_i, c)$ \\
    \midrule
    \texttt{fuse}$(i_o, i_i, i_f)$ & \rasep& Fuses two nested forall loops $i_o,i_i$ into one $i_f$. \\
    \multicolumn{2}{c}{$\ldots \forall_{i_o}\forall_{i_i} S$\hfill$ \xrightarrow[]{\textsf{fuse}(i_o, i_i, i_f)} $} & $\ldots \forall_{i_f} S~\textsf{s.t. fuse}(i_o, i_i, i_f)$ \\
    \midrule
    \texttt{reorder}$(i*)$            & \rasep&  Reorders the forall nodes based on a list of index variables $i*$. \\
    \multicolumn{2}{c}{$\ldots \forall_i \forall_{j\ldots} S$\hfill$ \xrightarrow[]{\textsf{reorder}(j,i,\ldots)}$}  & $\ldots \forall_j \forall_{i \ldots} S$ \\
    \bottomrule
  \end{tabu}
  \caption{Scheduling commands from TACO along with their CIN transformation~\cite{kjolstad2017tensor, kjolstad2019, senanayake2020}. $e[x'/x]$ denotes the expression $e$ with each occurrence of $x$ replaced by $x'$.}
  \label{table:scheduling-old}
\end{table}

\paragraph*{Format Language}
The format language proposed by Chou et al.~\cite{chou2018} decomposes a sparse tensor into per-dimension (or level) formats that each describes how to store the coordinates of one dimension of a tensor. As an example, the canonical compressed sparse row (CSR) compression format (see \Cref{fig:format-mem} for an example matrix) can be represented by an uncompressed (dense) dimension followed by a compressed (sparse) dimension. After the tensors have been described using level formats and scheduling transformations have been applied to the CIN, TACO generates code that iterates over the level formats of the expression.

\subsection{Capstan and Spatial}
\label{sec:rda}

RDAs improve performance and efficiency by removing overhead found in CPUs and GPUs. RDAs map programs in space, meaning multiple data elements are processed in the same clock cycle by pipelined and parallel compute units.

\begin{figure}
  \centering
  \includegraphics{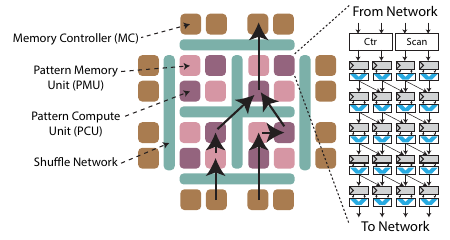}
  \caption{A high-level overview of the Capstan architecture, showing the opportunities for high-level parallelism across PCUs and vectorized parallelism within a PCU.}
  \label{fig:capstan}
\end{figure}

Capstan~\cite{rucker2021capstan}, shown in \autoref{fig:capstan}, is an RDA that targets sparse problems and is derived from Plasticine~\cite{plasticine}. A notable Capstan architectural contribution is a \textit{declarative-sparse model} for sparse iteration. 
This model divides sparse iterations into pattern headers and pattern bodies, where headers determine which (un)compressed iterations to run, and bodies use header iteration information to load, compute, and store data. An example of this paradigm for co-iteration of two compressed tensor levels can be found in \Cref{fig:declarative-sparse}.

Capstan is programmed with Spatial~\cite{spatial},\footnote{A full description of Spatial can be found at \url{spatial-lang.org}.} a parallel-pattern hardware domain-specific language for FPGAs and RDAs. 
Spatial's output is lowered to a streaming on-chip dataflow graph by the SARA compiler~\cite{zhang2021sara}, which handles low-level optimizations and inserts memory-consistency logic.

Spatial uses a map-reduce abstraction. 
Each \code{Foreach} or \code{Reduce} pattern is counter-indexed with an explicit parallelization factor; multiple levels of nested loops can be independently parallelized to exploit different program-level parallelism opportunities. 
Typically, the innermost loop is vectorized, and the outermost loop is replicated across pattern compute units (PCUs).
Capstan provides sparse iterator patterns---including  union and intersection combinations---in addition to dense ones.
Sparse patterns iterate by running on non-zero bit-vector elements using the index of the non-zero element instead of a counter.
Sparse patterns are key to Capstan's performance by enabling a declarative programming model (see \Cref{fig:scanners}).

Spatial has an explicit, decoupled, programmer-managed memory hierarchy. In a CPU, memory is managed using caches and demand misses; however, Spatial requires manually partitioning data into chunks that fit on-chip and controlling the corresponding data movement. 
Specifically, there are four programmer-controlled memory types, ranging from far to near: DRAM, SRAM, FIFOs, and registers, with the middle two mapping to Capstan's pattern memory units (PMUs).

\begin{figure*}
    \centering
    \footnotesize
    \begin{minipage}{0.39\textwidth}
          \subfloat[][Imperative code (similar to code generated by the TACO compiler) for SDDMM with compressed sparse row (CSR) A and B matrices. Numbers highlight assumptions inherent to the programming model]{%
        \label{fig:sddmm-imperative-code}
        \rule{0pt}{1.2in}%
\begin{lstlisting}[style=cppstyle]
(*\circone*)for (i = 0; i < C1_dim; i++) 
    for (jB = B2_pos[i]; jB < B2_pos[i+1]; jB++)
        j = B2_crd[jB](*\circtwo*)
        for (k = 0; k < D1_dim; k++)
          (*\circthree*)A[jB] (*\circfour*)+= B[jB]*C[i,k]*D[k,j]
\end{lstlisting}}
    \end{minipage} \hfill
    \begin{minipage}{0.6\textwidth}
    \centering
      \subfloat[][SDDMM mapped onto a streaming spatial programming model, in which the streamed values are elements of $B$. Gray blocks are memories partitioned into chunks and bound to physical memory units---either off-chip memory or on-chip scratchpads (PMUs). Yellow blocks bind computation spatially to the compute units (PCUs).]{
      \label{fig:sddmm-decl-model}
    \includegraphics[width=0.98\textwidth]{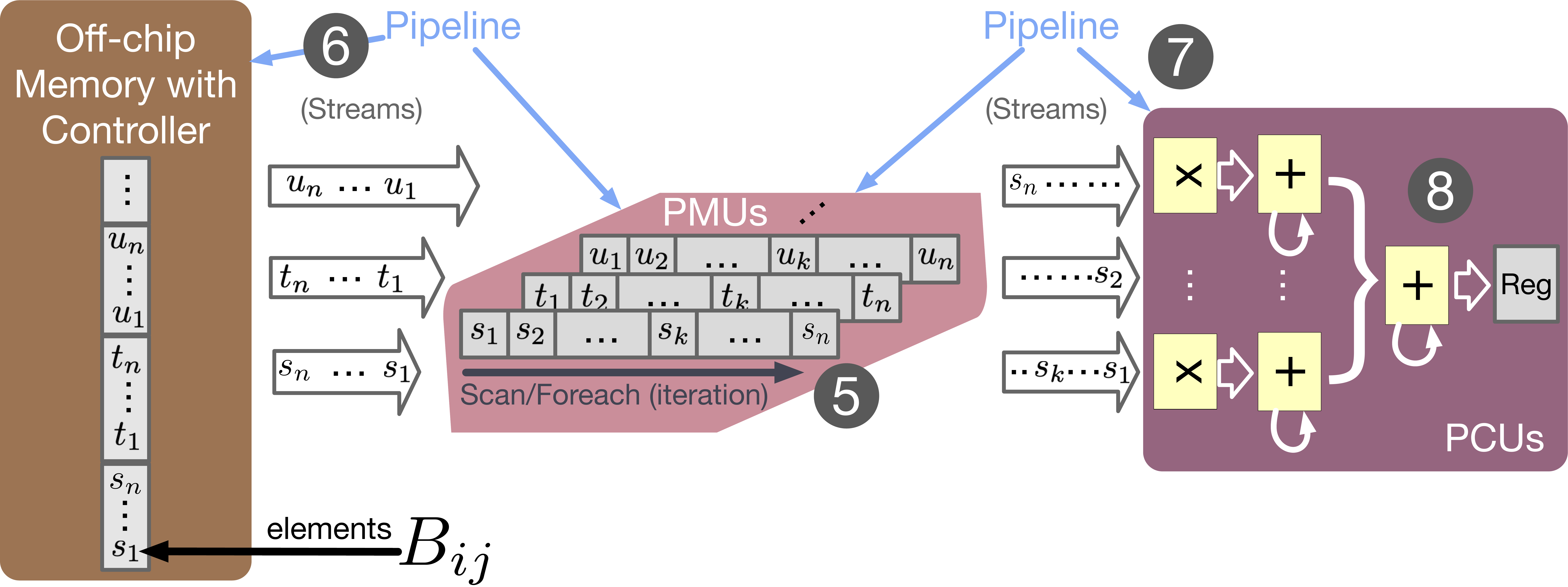}}
    \end{minipage}
    \caption{The SDDMM algorithm in an imperative programming model and as a spatial dataflow hardware configuration.}
\end{figure*}

\section{Running Example} 
To illustrate the gap between compiling sparse expressions in imperative code versus code generation for RDAs, we will introduce a running example.  
Consider SDDMM (sampled dense-dense matrix multiplication), a sparse linear algebra kernel used in machine learning. 
SDDMM defined in index notation is $A(i,j) = B(i,j) * C(i,k) * D(k,j)$ where A and B are sparse (and in this case CSR) matrices. 
However, index notation does not specify low-level control flow.
We can expand the index notation to the following imperative CIN:
\begin{align}
    \forall_i\forall_j\forall_k\left(A_{ij}\pluseq B_{ij} * C_{ik} * D_{kj}\right)
    \label{eqn:sddmm}
\end{align}
As shown in \Cref{fig:sddmm-imperative-code}, there is a clear path from this CIN statement to an imperative programming model, and many compiler decisions in prior work presuppose this programming model: \circone\ the $\forall$ nodes are directly converted into for-loops, \circtwo\ tensor accesses always load and store one element of the tensor at a time, \circthree\ tensor computation occurs only inside the innermost loop, and \circfour\ accumulations are implemented as temporally-repeated variable modifications.

To target a streaming spatial hardware backend, on the other hand, a compiler cannot leverage the above assumptions in the imperative programming model. Therefore, \name must address the following challenges when bridging the gap from an imperative to streaming spatial programming model: \circfive{}~the $\forall$ nodes must be traversed in a scanned fashion without temporal counters, \circsix\ data movement occurs between memory locations in vectorized streamed chunks parallelized across pipelines, \circseven\ tensor computation occurs when the data arrives as opposed to in the innermost loop, and \circeight\ parallel streams of computation are mapped in space to hardware resources.   
\Cref{fig:sddmm-decl-model} lays out the sparse SDDMM kernel as a streaming spatial program mapped onto the Capstan architecture, shown in \Cref{fig:capstan}, illustrating the compilation gaps that \name needs to address. The compiler targets the Capstan hardware by reorganizing the computation into a spatial dataflow program using a declarative, parallel-pattern (map-reduce) based functional programming abstraction with explicit management of fixed-size memories~\cite{spatial}.

\section{Scheduling Computation for Accelerators}
\label{sec:format-schedule}

A sparse tensor algebra expression can be computed using many different algorithms on different data structures, which motivates the separation of expressions, schedules, and data representation (formats)~\cite{kjolstad2020sparse,halide2012}. We introduce a new data representation property and scheduling language constructs. When combined with prior work on TACO formats~\cite{kjolstad2017tensor,chou2018} and scheduling constructs~\cite{kjolstad2019,senanayake2020}, they let end users reshape computation to expose sub-expressions that can be mapped to domain-specific accelerator patterns. Sparse domain-specific architectures (DSAs) are optimized for certain computation and iteration patterns dictated by the architectural decisions of the designer. Scheduling computation that efficiently (and correctly) maps to a hardware backend is key since these accelerator backends will have fixed resources, a fixed design, support for only certain features, and various tradeoffs. The scheduling transformation framework enables an end user, or an auto-scheduler, to exploit the hardware optimizations by reshaping the computation to expose sub-computation that will fit the target backend (with any other computation orchestrated by the host machine).

\name provides input languages that place tensors in explicit memory regions (to denote tensor scoping) and computation acceleration through specialized function mappings. We extend the data representation language of Chou et al.~\cite{chou2018} with a new memory placement property that lets the user (or a future auto-scheduler) explicitly place tensor data structures in either \textit{off-chip (global)} or \textit{on-chip (local)} memory. Furthermore, we also extend the scheduling language framework presented by Senanayake et al.~\cite{senanayake2020} with three new scheduling commands that map kernels to optimized, configured backend implementations.
Finally, we demonstrate how a user can leverage the input languages to reshape computation to Capstan's parallel patterns.\footnote{Although we target the Capstan parallel patterns in this work, the proposed scheduling construct can in future work be adapted to target other hardware accelerators or even hand-optimized library calls.}

Thus, we raise the abstraction of the low-level hardware to the high-level tensor input languages of \name and separate out the scheduling commands that map the computation to specialized hardware backends, which can be used by a hardware expert or by a future auto-scheduler.

\begin{figure}[t] 
    \footnotesize
    \centering
    \begin{lstlisting}[style=cppstyle]
// Define off-chip (global) tensor formats
Format csr_off({uncompressed,compressed}, offChip);
Format rm_off({uncompressed,uncompressed}, offChip); 
Format cm_off({uncompressed,uncompressed}, {1,0}, offChip);
// Declare input and output tensors
Tensor<int> A({N,N}, csr_off);
Tensor<int> B({N,N}, csr_off);
Tensor<int> C({N}, rm_off);
Tensor<int> D({N}, cm_off);

// Define SDDMM computation (algorithm).
IndexVar i, j, k;
A(i, j) = B(i, j) * C(i, k) * D(k, j);

// Scheduling language: Define environment variables
IndexStmt stmt = A.getAssignment();
stmt = stmt.environment(innerPar, 16);
stmt = stmt.environment(outerPar, 2);
// Precompute accumulation into a ws register 
// to accelerate it using a Reduce pattern
Tensor<int> ws(on);
stmt = stmt.precompute(B(i,j)*C(i,k)*D(k,j), {}, {}, ws);
stmt = stmt.accelerate(forall(k, ws+=B(i,j)*C(i,k)*D(k,j)),
                Spatial, Reduction, innerPar);
                
// Perform computation by generating and executing code
std::cout << A << std::endl;
    \end{lstlisting}
    \caption{\label{fig:sddmm-user}
        Abridged C++ \name input code for SDDMM.
    }
\end{figure}

\subsection{Format Language: Explicit Tensor Memory Regions}
\label{sec:coarse-mem}

The data representation API of \name lets a user place a tensor into their memory region of choice, which in our system is either off-chip memory or in the memory of the accelerator.\footnote{Users only state whether to place memory on the accelerator or not. Our accelerator, however, may have many different types of memories on-chip and our automated memory analysis determines which specific accelerator memory to use.} The off-chip memory region is globally accessible to all backends involved in the computation (host and accelerators) whereas on-chip tensors are only locally accessible locally to one backend. This memory region annotation within the format language is essential in denoting the scope in which specific tensor data lives (within a memory hierarchy), describing where tensor data should be initialized, and reasoning about what actions need to occur (like data transfers) to get the data in scope before computing on those tensor values.

With the explicit memory region annotation, information to transfer input data to the accelerator (on-chip) or results back to the host (off-chip) is encoded within the CIN representation as an assignment statement. Moreover, the \code{precompute} scheduling command, together with the data representation API, lets users create new temporary tensors to store intermediate values and place them in accelerator memories. Our SDDMM example demonstrating the on- and off-chip communication API is shown in \Cref{fig:sddmm-user} (with the final Spatial code generated by \name in \Cref{fig:sddmm-spatial}). \Cref{fig:sddmm-onchip} then demonstrates the use of the \code{precompute} scheduling command to transfer data between different memory regions by inserting temporary tensor variables with a different memory region. We require the explicit on- versus off-chip placement (via the scheduling language) because input tensors could start out in various memory locations and host$\leftrightarrow$accelerator transfers could happen at multiple program locations. 
\Cref{fig:sddmm-onchip} shows two schedules, along with their CIN, that use different memory regions. Separate on-chip and off-chip memory regions also let our compiler represent tensor partitioning (like blocking or tiling) on the host processor, with faster accelerated computation on-chip in the DSA.

\begin{figure}
\footnotesize
\begin{minipage}{\linewidth}
  \subfloat[][New CIN after partial on-chip loads of $C_{rows}$ and $D_{cols}$ in the $j$-loop body using two \texttt{precompute} commands.]{
    \begin{minipage}{\linewidth}
    \centering
    \begin{tabular}{c}
\begin{lstlisting}[style=cppstyle]
stmt = stmt.precompute(C(i,k), {k}, {k}, C_on)
stmt = stmt.precompute(D(k,j), {k}, {k}, D_on)
\end{lstlisting} 
  \end{tabular}
    \begin{align*} 
    \forall_i \forall_j(&\forall_k (A_{ij} \mathrel{+}= B_{ij} * C^\text{on}_k * D^\text{on}_k) \\
     & \textsf{ where } \forall_k( C^\text{on}_k = C_{ik}) \\
     & \textsf{ where } \forall_k(D^\text{on}_k = D_{kj}))
    \end{align*}
    \end{minipage}
  }
 \end{minipage} \hfill
 \begin{minipage}{\linewidth}
  \subfloat[][New CIN after initial load of $C$ and $D$ entirely before computation loops using two different \texttt{precompute} commands.]{
    \begin{minipage}{\linewidth}
    \centering
    \begin{tabular}{c}
\begin{lstlisting}[style=cppstyle]
stmt = stmt.precompute(C(i,k), {i,k}, {i,k}, C_on)
stmt = stmt.precompute(D(k,j), {k,j}, {k,j}, D_on)
\end{lstlisting}
   \end{tabular}
     \vfill
    \begin{align*} 
    \forall_i \forall_j \forall_k (&A_{ij} \mathrel{+}= B_{ij} * C^\text{on}_{ik} * D^\text{on}_{kj}) \\
     & \textsf{ where } \forall_i\forall_k( C^\text{on}_{ik} = C_{ik}) \\
     & \textsf{ where } \forall_j\forall_k(D^\text{on}_{kj} = D_{kj})
    \end{align*}
    \end{minipage}
  }
\end{minipage}
 \caption{ \label{fig:sddmm-onchip}
 Two SDDMM schedules and their corresponding CIN statements demonstrating the mapping of formats on- and off-chip for accelerator backends (in contrast to the default schedule in \cref{eqn:sddmm}). Tensors $A, B, C^\text{on}, D^\text{on}$ have on-chip formats and $C,D$ have off-chip formats.}
\end{figure}

\subsection{Scheduling Language: Mapping Sub-computation}
\label{sec:schedule-map}

Through \name, users leverage schedules to reshape CIN sub-statements to expose sub-computations that can be mapped to high-level accelerated primitives, which in this paper are the Spatial parallel patterns. Given any CIN statement $S$ that includes a sub-statement $S'$ where $S \defeq \ldots S' \ldots$, and $S'$ has an equivalent instruction $f$ for a given platform, the scheduling language can transform $S$ such that $S'$ is isolated. Then, the sub-statement $S'$ can be replaced and computed using the specialized pattern or function $f$ for a given backend instead of being lowered directly to code. 

Concretely, consider the simple vector-vector multiplication statement $S~\defeq~S'~\defeq~(\forall_i a_i=b_i*c_i)$ where all vectors are currently off-chip. Assuming there exists a vectorized multiplier block $f_\text{mul}$(out, $\text{in}_1$, $\text{in}_2$) for a given backend, the \code{map} command can be used in conjunction with the \code{precompute} command to optimize the kernel. The transformation is demonstrated in equations 2--4:

  {
\begin{align}
    S' \defeq (\forall_i a_i = b_i * c_i)  & \xrightarrow[]{\textsf{precompute(}
    b_i*c_i, \{i\}, \{i\}, a^\text{on})} S''
\end{align}}
The above precompute transformation results in a new statement where the vector-vector multiplication result is stored into an on-chip result $a^\text{on}$ and then the on-chip result is stored back off-chip into $a$.

Then, another precompute transformation is called for each off-chip input tensor so that the vector-vector multiplication is computed using only on-chip inputs
\begin{equation}
  \arraycolsep=2pt
  S'' \defeq \left\{
  \begin{array}{ll}
    \forall_i a_i = a^\text{on}_i & 
     \xrightarrow[]{\forall t \in \{b,c\})~\textsf{precompute(}
    t_i, \{i\}, \{i\}, t^\text{on})} S''' \\ 
     \multicolumn{2}{l}{\textsf{where}~\forall_i a^\text{on}_i = b_i * c_i}
\end{array}
\right.
\end{equation}
where $t^\text{on}$ denotes an on-chip tensor format and $t$ denotes an off-chip format for all $t \in \text{tensors}(e)$. 

Lastly, the vector-vector multiplication kernel is mapped to the vectorized multiplier $f_\text{mul}$ using only on-chip tensors as operands $f_\text{mul}(a^\text{on}, b^\text{on}, c^\text{on})$. 
\begin{equation}
{\footnotesize
  \arraycolsep=2pt
  S''' \defeq \left\{
\begin{array}{rllr}
  \forall_i a_i = a^\text{on}_i&\multicolumn{2}{c}{\xrightarrow[]{\textsf{map}(\forall_i a^\text{on}_i = b^\text{on}_i * c^\text{on}_i, \text{backend}  , f_\text{mul})}} & \forall_i a_i = a^\text{on}_i\\    
    \multicolumn{2}{l}{\textsf{where}~\forall_i a^\text{on}_i = b^\text{on}_i * c^\text{on}_i} & \multicolumn{2}{r}{\textsf{where}~f_\text{mul}(a^\text{on}, b^\text{on}, c^\text{on})} \\
     \multicolumn{2}{l}{\textsf{where}~\forall_i b^\text{on}_i = b_i}&\multicolumn{2}{r}{\textsf{~s.t. map}(\text{backend}, f_\text{mul})} \\
     \multicolumn{2}{l}{\textsf{where}~\forall_i c^\text{on}_i = c_i}&\multicolumn{2}{r}{\textsf{where}~\forall_i b^\text{on}_i = b_i} \\
     && \multicolumn{2}{r}{\textsf{where}~\forall_i c^\text{on}_i = c_i}
     \end{array}
     \right.
     }
\end{equation}
We also introduce define a new \code{accelerate} scheduling command that combines all of these intermediate steps. This command is a compound command consisting of one or more \code{precompute} commands and a \code{map} command, and is necessary to map any sub-statement to a new backend function. 

Given that $S~\defeq~\ldots S'\ldots$ and $S'~\defeq~\forall_{i*} a = e$, we define the \code{accelerate} transformation below.

\begin{equation}
  S \xrightarrow[]{\textsf{accelerate}(S', \text{backend}, f, c)} S_\text{new}\label{eqn:accelerate}
\end{equation}

is equivalent to

\begin{equation}
  \arraycolsep=2pt
  \begin{array}{rll}
    \ldots S' \ldots &\xrightarrow[]{\textsf{precompute(}e, i*, i*, a^\text{on})}  \\ 
    & \xrightarrow[]{\text{For all}~t \in \text{tensors}(e))~\textsf{precompute(}t_i, i*, i*, t^\text{on})} \\
    &\xrightarrow[]{\textsf{map}(S'[t^\text{on}/t~\text{for all}~t \in \text{tensors}(S')], \text{backend}, f, c)} S_\text{new}
\end{array}
\end{equation}

Intuitively, the \code{accelerate} command first precomputes all off-chip tensors on-chip for a the sub-statement that is being accelerated and then maps the on-chip tensors to the backend function $f$ for substituted computation. 

We also describe a specific use-case for the \code{accelerate} command in our implementation targeting Spatial. This command is used to accelerate accumulations by calling Spatial's \code{Reduce} pattern, which maps to Capstan's optimized reduction tree hardware within a PCU. \name leverages the \code{accelerate} command to transform forall-with-accumulation patterns into a \code{Reduce} pattern (\Cref{fig:sddmm-user}, lines 28--30). 
More specifically, the \code{accelerate} command may be used with the \code{precompute} scheduling command to transform an accumulation loop $\forall(\ldots \pluseq \ldots)$ into a CIN sub-statement with the following \code{forall} pattern: ${\forall(\ldots=\text{ws}~\textsf{where}~\text{ws}\mathrel{+}=\ldots)}$, where ws is scalar. The transformed $\forall(\ldots)$ statement is accelerated by replacing the CIN statement with a Reduce parallel pattern using the \code{accelerate} command (which under the hood calls the \code{map} command). Our compiler also has optimization passes that can automatically \code{accelerate} common CIN sub-statements to more optimized implementations. One such automatic scheduling pass detects CIN sub-statements that loop over an array transferring a single element of data at a time, $\forall(i, t_1(i) = t_2(i))$, and maps them to bulk memory load or store functions.

Finally, metadata must be passed directly to the global scope in order to configure the backends that call their specialized functions $f$ when using the \code{map} and \code{accelerate} commands. Therefore, we introduce an \code{environment} command to set these metadata configuration variables to values. Lines {44--45} in \Cref{fig:sddmm-user} shows the configuration of both the inner and outer parallelization factors to pass to the Spatial compiler~\cite{spatial}. With the \code{environment} command, users can use our compiler to search the design space of kernels parameterized by these metadata values. Specifically, an end-programmer can now perform design-space exploration of the backend hardware schedules and tensor-algebra kernels using high-level index notation and scheduling transformations (including global configuration parameters) without direct knowledge of the backend architecture or its programming model. 

Although transformations of our SDDMM example in \Cref{fig:sddmm-user} are very specific to parallel patterns, it demonstrates the importance of the input languages of \name in order to generate valid algorithms for any backend or pattern. The transformation framework can more generally be applied to any CIN statement and backend combination. A full list of scheduling commands needed to target the hardware from \name, and their corresponding descriptions, can be found in \Cref{table:scheduling-old} and \Cref{table:scheduling-new}. 

\begin{table}
  \centering
   \tabulinesep=3pt
  \footnotesize
     \begin{tabu} to \linewidth {@{}p{1.2in}@{}c@{\,\,}>{\scriptsize}X@{}}
    \toprule
    \rowfont{\footnotesize\sffamily\bfseries} Scheduling Commands & & Description\\
    \midrule
    \hangindent=1em\texttt{map}$(S, \text{backend}, f, c)$ & \rasep & Maps a CIN statement $S$ to a backend-specific computation strategy (specialized block, function, pattern, or instruction) $f$ with some optional constant factor, $c$. \\
    \multicolumn{2}{c}{$\ldots S$\hfill$ \xrightarrow[]{\textsf{map}(
    S, \text{backend}, f, c)}$}  & $\ldots f(\text{tensors}(S), V^{\dagger}, c)~\textsf{s.t. map}(\text{backend}, f)$ \\
    \multicolumn{3}{p{0.9\linewidth}}{Where $V$ is the set of variables $\{i*, r*, \text{var}*\}$ defined by the scope of the 
    CIN sub-tree right before the statement $S$.} \\
    \midrule
       \hangindent=1em\texttt{accelerate} $(S, \text{backend}, f, c)$ &\rasep & A compound scheduling command that accelerates a sub-statement $S$ by precomputing all tensors of $S$ into on-chip tensors for a new expression $S'$ and then maps an equivalent $f$ onto $S'$ (the new statement with all on-chip tensors). \\
    \multicolumn{3}{c}{See \cref{eqn:accelerate} for the formal transformation definition.}  \\
    \midrule
    \texttt{environment}$(\text{var}, c)$ &\rasep & Sets a global hardware configuration variable to some value, $c$. \\
    \multicolumn{2}{c}{$S$\hfill$ \xrightarrow[]{\textsf{environment}(\text{var}, c)}$}  & $S~\textsf{s.t.}~\text{var}=c$ \\
    \bottomrule
  \end{tabu}
  \caption{New scheduling commands necessary to target DSAs.}
  \label{table:scheduling-new}
\end{table}

\section{Memory Analysis}
\label{sec:mem-analysis}

When compiling to accelerators, the compiler will often need to generate code to manage explicit fixed-length memories. 
Hardware accelerators (including GPUs) typically have memory hierarchies with multiple physical memory types that are explicitly managed by software, also called explicit, decoupled data orchestration~\cite{pellauer2019,rucker2021capstan,plasticine,mei2015gpu,dadu2019towards,hegde2019extensor}. 
These memory types often have different capacities, transfer speeds, access patterns, and scopes (notably, only some are host-visible); the compiler must bind data structures to memories and generate transfers between them. 
The different sub-arrays involved in sparse tensors---coordinates, positions, and values---are also often mapped separately based on a combination of the sub-array and memory properties, complicating the analysis. 

\name analyzes the memory needs of format-annotated CIN to generate Spatial kernels for Capstan.
The format abstraction gives end users explicit control over tensor placement for one level of memory hierarchy, which we call \textit{coarse-grained} memory pinning. 

The compiler then automatically binds tensor format sub-arrays (positions, values, etc.) to Spatial memory types (which later become physical Capstan memory) using \textit{fine-grained} memory pinning. The automatic inference analysis is crucial in avoiding unwieldy end-user memory management, which would require decisions on the Cartesian combination of: the number of input tensors for a kernel, levels per tensor, sub-arrays per level format, and the memory analysis for each sub-array (which includes memory type matching, memory allocation, and multiple data transfer locations).
Finally, CIN assignment nodes are lowered to inter-region data transfers (e.g., DRAM to SRAM loads and vice versa).
This analysis ensures that memory properties and preconditions are satisfied, ensuring valid placement of tensor arrays.

\subsection{Fine-Grained Array Memory Inference}
\label{sec:fine-mem}

\name emits tensor sub-array allocations and memory transfers during top-down CIN traversal, accounting for memory access patterns and physical properties. 
Fine-grained memory binding addresses multiple issues: 1) which physical memory type to allocate for a given tensor array, 2) where (in which pattern body) to allocate the chosen physical memory, and 3) where (in which pattern body) should our compiler emit inter-memory transfers. Incorrect analysis---incompatible memory allocations, late allocations, and missed data transfers---will cause hardware simulation errors or invalid kernel computations. 
The memory binding analysis does not consider array sizes since it allocates the maximal possible size for one unit of memory; it assumes every array will fit in this size constraint based on computation tiling to the accelerator as described by the scheduling language.

\name ensures the following preconditions when binding format sub-arrays of tensors to physical memory types: 

\noindent \textbf{Dense DRAMs.} Arrays of every off-chip tensor (input and output) are placed in dense DRAM initialized by the host. 

\noindent \textbf{Sparse DRAMs.} These provide an interface for direct off-chip random accesses of sparse data. 
This is useful when there is no identifiable working set to bring on-chip.
    
\noindent \textbf{Dense SRAMs.} The system will only bind arrays with affine access patterns to dense SRAMs, including position arrays (addressed in an $addr, addr+1$ fashion) and values arrays of fully dense formats (generally traversed linearly).

\noindent \textbf{Sparse SRAMs.} The system binds any on-chip, small, fixed-size arrays that have an access pattern with reuse but random accesses to sparse SRAMs.

\begin{figure}
    \centering
    \includegraphics[width=\linewidth]{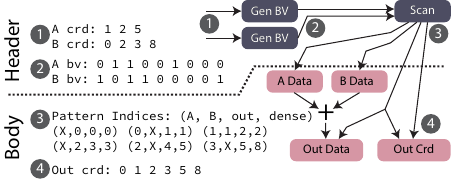}
    \caption{
        Element-wise co-iteration of two compressed vectors in Capstan. A sparse bit-vector scanner divides the problem into parallel map iterations, followed by memory atomics and reductions (pink blocks denote physical memories). 
        \label{fig:declarative-sparse}
    }
\end{figure}

\noindent \textbf{Bit Vectors.} Bit vectors are special on-chip integer streams (\Cref{fig:declarative-sparse}) that hold compressed coordinate information. 
\name automatically generates and manages bit vectors whenever a compressed-compressed co-iteration occurs.

\noindent \textbf{FIFO Buffers.} Arrays accessed linearly with certain access patterns may be bound to (seemingly infinite) FIFO buffers. 
The compiler cannot enqueue excess data that is not popped, cannot pop data before its storage lifetime ends, and must pop the data precisely when the storage lifetime ends. 
This restricts FIFOs to coordinate arrays of sparse tensors and value arrays that are accessed in order. 

\noindent \textbf{Registers.} On-chip scalar variables are bound to registers. 

\subsection{Data Transfer Analysis}

After the compiler has chosen an accelerator memory type for a tensor, it inserts code to allocate memory and to transfer data to and from the memory. Tensor arrays are allocated at the loop level just above their first use, for ease of analysis, but hoisting them and inserting reset code between iterations is also possible.

\begin{figure}
    \footnotesize
    \centering
    \includegraphics[width=\linewidth]{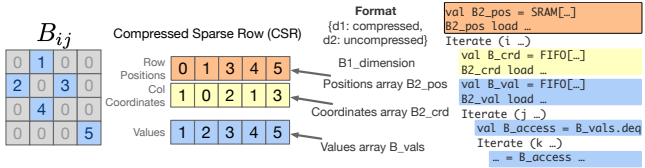}
    \caption{
    Example sparse $B$ matrix used in SDDMM with its corresponding data structure and format arrays. The right details Spatial pseudocode generated by our work, demonstrating array memory allocations, accesses, and data transfers interleaved into the computation.  
    }
    \label{fig:format-mem}
\end{figure}

Sparse tensors have multiple index arrays and a value array whose data must be transferred to and/or from the accelerator. 
Their value arrays store the actual values, while each of its compressed level stores the compressed indices in two arrays: positions and coordinates. 
Consider our running SDDMM example (illustrated in \Cref{fig:format-mem}), where $B$'s CSR format list (an uncompressed/dense dimension followed by a compressed dimension) describes the access $B_{ij}$, as well as the iteration over the index variables \{i, j\} and the loops $\forall_i\forall_j$. The innermost level of $B$ corresponds to the index variable $j$ and iterates over the position and coordinate arrays \code{B2\_pos} and \code{B2\_crd}, respectively. This means both \code{B\_vals} and \code{B2\_crd} elements will be accessed inside the $j$-loop, so both arrays must be allocated right before in the $i$-loop body, with \code{B2\_pos} being accessed in the $i$-loop and allocated at the top.

Our lowering algorithm ensures tensors are filled with values before their use by making the following assumptions: 1) value-array elements are always accessed at the corresponding loop body of the innermost mode of a tensor, 2) coordinate array elements are always accessed at the index loop body corresponding to that mode, and 3) position array elements are always accessed one loop higher. If a higher loop does not exist, the arrays are accessed at the start of the kernel.
All allocations occur before the loop header of the current loop (also the loop body of the preceding loop-nest), and all data transfers occur immediately after their associated allocations.

Our lowering algorithm must access value elements at the loop body of a tensor's innermost mode and not lower in the loop-nest hierarchy. 
We may not assume that array elements are accessible anywhere after array declaration because this is not possible for memories that do not support random access. Using a FIFO for an in-order traversal of the modes, for example, requires that the FIFO values be accessed precisely at the level of the last tensor mode and only used for one iteration of that loop. \Cref{fig:format-mem} demonstrates that if the value array of $B_{ij}$ was bound to a FIFO with the iteration pattern of the computation as $\forall_i\forall_j\forall_k$, the \code{B\_vals} array elements would have to be accessed in the $j$-loop (corresponding to $B$'s last mode) instead of the $k$-loop (the innermost loop) as in the CPU case.

With our memory analysis, we simplify end-user programming. 
Users decide only whether the tensor is on-chip or off-chip, while our compiler manages the fine-grained arrays associated with each compressed tensors. 
Our algorithm design also lets \name perform lowering transformations and optimizations at a fine-grained (per-array) level.

\section{Co-iteration Lowering as Rewrite Rules}
\label{sec:coiteration-rewrite}

\begin{figure}
    \footnotesize
    \centering
    \begin{lstlisting}[style=spatial]
// Uncompressed iteration and reduction
Foreach(len by inc par p) {i_dense => ...}
Reduce(reg)(len by inc par p) {i_dense => ...}
MemReduce(mem par mp)(len by inc par p) {i_dense => ...}
// Compressed single iteration (Reduce not shown)
Foreach(len by inc par p) {pos => ...}
Foreach(Scan(par=p, len=l, bitvector_A.deq)) 
    {A, i_dense => ...}
// Compressed-compressed coiteration (Reduce not shown)
Foreach(Scan(par=p, len=l, bitvector_A.deq, 
        bitvector_b.deq)) {A, B, out, i_dense => ...}
    \end{lstlisting}
    \caption{
        Spatial parallel patterns for compressed (sparse) and uncompressed (dense) mode-level iterations. 
    }
    \label{fig:scanners}
\end{figure}

In many situations, DSAs and accelerators accelerate computation using streams of data.
These streams flow through functional hardware blocks that produce new output streams. 
Because the blocks are functional mappings from input to output streams, general-purpose control flow statements like conditionals are no longer available, TACO's lowering machinery can no longer be used.

\subsection{Co-iteration Rewrite System}
Sparse accelerators further speed up sparse tensor computations by contracting together tensor elements efficiently, and a good compiler must support these algorithms natively.
For example, DSAs may have specialized hardware blocks with clever algorithms and optimized data structures to increase performance, such as the examples in \Cref{fig:scanners}. 
One such data structure is the bit-vector format, which densely packs sparse coordinate information~\cite{rucker2021capstan,briggs1993efficient}. 
The bit-vector format accelerates kernels by enabling hierarchical intersections using bit-vector trees, uncompressed blocked iteration of sparse data, and efficient tensor contractions using boolean logic. 

\name introduces a rewrite system for matching fused iteration patterns to a corresponding backend behavior, which is shown in \Cref{tab:rewrite-rules}. 
The lowering mechanism traverses the CIN IR recursively. 
When a CIN \code{forall} node is found, the lowerer decomposes the level's fused tensor iterator contraction set, $I = \mT_1 \circ \mT_2 \circ \cdots \circ \mT_n~\textsf{s.t.}~\circ \in \{\cup, \cap\}, n \geq 1$, into smaller tensor iterator contraction subsets based on the iterator formats for that level and the contraction type. 
The tensor iterator contraction subsets generated are those supported by the backend. 
For example, if a backend has a custom instruction, \code{or-and}, to compute $(a \cup b) \cap c$ given three compressed inputs, its rewrite rule would be: 
\texttt{\iterFn}$[(\mathcal{C}_1 \cup \mathcal{C}_2)
\cap \mathcal{C}_3]~\Rightarrow~$\textbf{emit} \code{or-and}. 
If a backend does not have a rewrite rules (e.g., intersection but not union support), the rewriter will map unmatched computations to the host.

\newcommand{\Rasep}{{\hskip-1ex$\Rightarrow$}}
 \begin{figure}
  \centering
   \tabulinesep=-1pt
   \newcommand{\secsep}{\midrule}
  \footnotesize
       \centering
       \fbox{
   \begin{minipage}{0.8\linewidth}
    Let the set of iterator contractions for a given $\forall$ node be:
    $$I = \mT_1 \circ \mT_2 \circ \cdots \circ \mT_n~\textsf{s.t.}~\circ \in \{\cup, \cap\}, n \geq 1$$
     Let format($I$) = format($\mT_1) \circ \text{format}(\mT_2) \circ \ldots \circ \text{format}(\mT_n)$
     format($\mT_n$) = $\mathcal{C}_n$ if $\mT_n$ is compressed, format($\mT_n$) = $\mathcal{B}_n$ if $\mT_n$ is a bit vector,
     format($\mT_n$) = $\mathbb{U}$ (the universe) otherwise.
   \end{minipage}}
   \vskip1ex
   \begin{tabu} to \columnwidth {@{}c@{\hskip2ex}p{1.05in}@{}c@{\,\,}>{\scriptsize}X@{}}
    \toprule
     \multicolumn{2}{l}{\texttt{\iterFn}[$\text{format}(I)$]} & \Rasep & \textbf{emit} <backend block behavior> \\
    \midrule
     
     \multirow{6}{*}{\rotatebox{90}{\sffamily\bfseries Single-Iteration}} 
     & \texttt{\iterFn}[$\mathbb{U}$] & \Rasep & \emititerationcolor{\denseIter{}} \\
     &\texttt{\iterFn}[$\mathcal{B}_1$] & \Rasep & \emititerationcolor{\genBVResult{}} \\
     && & \emititerationcolor{\sparseBVIter{}} \\
     &\texttt{\iterFn}[$\mathcal{C}_1$ \textbf{and} & \Rasep & \emititerationcolor{\genBVone} \\
     &\hspace*{13mm} $\mT_1$ is result] & & \texttt{\iterFn}($\mathcal{B}_1$) \\
     &\texttt{\iterFn}[$\mathcal{C}_1$] & \Rasep & \emititerationcolor{\sparseBVIter{}} \\\secsep

     \multirow{3}{*}{\rotatebox{90}{\sffamily\bfseries Universe}} 
     &\texttt{\iterFn}[$\mathbb{U} \cup \_~]$ & \Rasep & \texttt{\iterFn}($\mathbb{U}$) \\
     &\texttt{\iterFn}[$~\_ \cup \mathbb{U}$] & \Rasep & \texttt{\iterFn}($\mathbb{U}$) \\
     &\texttt{\iterFn}[$\mathbb{U} \cap \mathbb{U}$] & \Rasep & \texttt{\iterFn}($\mathbb{U}$) \\\secsep

     \multirow{2}{*}{\rotatebox{90}{\sffamily\bfseries Comp.}} 
     &\texttt{\iterFn}[$\mathcal{C}_1 \cap \mathbb{U}$] & \Rasep & \texttt{\iterFn}($\mathcal{C}_1$) \\ 
     &\texttt{\iterFn}[$\mathbb{U} \cap \mathcal{C}_2$] & \Rasep & \texttt{\iterFn}($\mathcal{C}_2$) \\\secsep

     \multirow{6}{*}{\rotatebox{90}{\sffamily\bfseries Co-Iteration}} 
     &\texttt{\iterFn}[$\mathcal{C}_1 \circ \mathcal{C}_2$] & \Rasep & \emititerationcolor{\genBVone{}} \\
     && & \emititerationcolor{\genBVtwo} \\
     && & \texttt{\iterFn}($\mathcal{B}_1 \circ \mathcal{B}_2$) \\
     &\texttt{\iterFn}[$\mathcal{B}_1 \circ \mathcal{B}_2$] & \Rasep & \emititerationcolor{\genBVResult{}} \\
     && & $\circ=\cup\Rightarrow$\,\emititerationcolor{\interIter{}}\\ 
     && & $\circ=\cap\Rightarrow$\,\emititerationcolor{\unionIter{}}\hskip-1ex\\ \secsep

     \multirow{2}{*}{\rotatebox{90}{\sffamily\bfseries Base}}
     &\texttt{\iterFn}[ $\_$ ] & \Rasep & format($\mT_{1k})=\texttt{\iterFn}(\mT_1 \circ \ldots \circ \mT_k$), largest $k \leq n$ that produces a match \\
     && & \texttt{\iterFn}(format($\mT_{1k} \circ ... \circ \mT_n$)) \\ 
    \bottomrule
  \end{tabu}
  \caption{General rewrite system that lowers combined tensor iteration and contractions to declarative constructs (found in \Cref{fig:scanners}), called by the \textproc{lower} function which recursively traverses the CIN IR and emits low-level IR.
        Specifically, this algorithm lowers $\forall$ nodes to the declarative-sparse programming model of sparse iteration theory. Blue statements have side effects that emit low-level IR code.
        }
  \label{tab:rewrite-rules}
\end{figure}

\subsection{Scheduled CIN Lowering Implementation}
Lowering from a scheduled CIN statement to Spatial parallel patterns is straightforward. 
Environment variables set by the schedule are emitted first so that they are globally scoped. 
Next, \name recursively traverses the CIN and replaces \code{map}-scheduled statements with optimized hardware functions or instructions.
To demonstrate, the example on line 34 in \Cref{fig:sddmm-user} would cause the lowerer to emit a \code{Reduce} pattern. 
\name then automatically lowers remaining $\forall$ nodes (those not scheduled with a \code{map}) to the \textit{declarative-sparse programming model} (see \cref{tab:rewrite-rules}). 
Instead of mutable counters, the compiler determines the iteration space (min until max by step) and uses pattern indices within the body to access tensor data.

When needed, elements of one to two input tensor coordinate arrays are streamed into a packed bit vector, as shown in \Cref{fig:declarative-sparse}. 
The packed bit vector stores compressed coordinate information using a $1$ where a coordinate is nonzero and a $0$ at all other positions. 
Two compressed bit vectors are then either logically AND-ed (for intersection/multiplication) or OR-ed (for union/addition) by the sparse bit-vector scanner, depending on the tensor algebra operator. 
The system then emits two scanner loops: one calculates position sub-array entries by counting the number of nonzero results, and the other computes entries for the value sub-array.

Finally, \textproc{lower} recurses on the computation in the body of the second scanner, using atomic accesses to sparse SRAMs for future value-array computation.
This lowerer implementation eliminates the need for temporal loads/stores, conditional control flow (if-statements and while-loops), and tensor union decomposition (into unions of disjoint intersections). \Cref{fig:sddmm-spatial} shows the final Spatial code outputted by \name for our SDDMM running example.

\begin{figure}[t] 
    \centering
    \begin{lstlisting}[style=spatial,style=tight]
// Spatial header code
...

// Initialize all DRAM arrays
val A2_pos_dram = DRAM[T](nnz_max)
...

Accel {
  val B2_pos = SRAM[T](nnz_accel_max)
  B2_pos load B2_pos_dram(0::(B1_dim + 1) par ip)
  Foreach (C1_dim by 1 par bp) { i =>
    val A_vals = FIFO[T](16)
    val A2_crd = FIFO[T](16)
    val A2_pos = SRAM[T](nnz_accel_max)
      
    val jB_start = B2_pos(i)
    val jB_end = B2_pos((i + 1))
    val jB_len = jB_end - jB_start

    val B2_crd = FIFO[T](16)
    B2_crd load B2_crd_dram(jB_start::jB_end par 1)
    val B_vals = FIFO[T](16)
    B_vals load B_vals_dram(jB_start::jB_end par 1)

    Foreach (jB_len by 1 par 1) { jB =>
      val j = B2_crd.deq
      val B_hoisted = B_vals.deq
        
      val D_vals = SRAM[T]((nnz_accel_max / 4))
      D_vals load D_vals_dram((j*D1_dim)::((j+1)*D1_dim) par ip)
      val C_vals = SRAM[T]((nnz_accel_max / 4))
      C_vals load C_vals_dram((i*C2_dim)::((i+1)*C2_dim) par ip)
        
      val tjA_vals = Reg[T](0.to[T])
      Reduce(tjA_vals)(D1_dim by 1 par ip) { k => 
        ((B_hoisted * C_vals(k)) * D_vals(k))
      } { _ + _ }
      A_vals.enq(tjA_vals)
      A2_crd.enq(j)
    }
    A2_pos(i + 1) = jB_end
    A_vals_dram stream_store_vec(jB_start, A_vals, jB_len)
  }
}
    \end{lstlisting}
    \caption{\label{fig:sddmm-spatial}
        SDDMM Spatial code generated by \name. 
    }
\end{figure}

\section{Evaluation}
We demonstrate significant performance improvements when using \name to target an RDA over compiling to a CPU or GPU, showing the the usefulness of our compiler from the perspective of a data analyst or computational end user interested in a fast sparse tensor algebra library. And from the perspective of a performance/hardware engineer tasked with utilizing a sparse accelerator, our results show that \name provides increased productivity compared to writing the applications in a low-level language at the abstraction of the hardware.

\label{sec:evaluation}

\subsection{Methodology}
\label{sec:methodology}

\begin{table}
  \centering
  \footnotesize
  \begin{tabu}{llrr}
    \toprule
    \rowfont{\sffamily\bfseries} & & \multicolumn{2}{c}{Lines of Code} \\\cmidrule(lr){3-4}
    \rowfont{\sffamily\bfseries} Name & Expression & \multicolumn{1}{c}{Input}& \multicolumn{1}{c}{Spatial}\\
    \midrule
    \sffamily\bfseries SpMV        & $y_{i} = \sum_j A_{ij}x_j$                       & 10 & 44 \\
    \sffamily\bfseries Plus3       & $A_{ij} = B_{ij} + C_{ij}+ D_{ij}$               & 8 & 91 \\
    \sffamily\bfseries SDDMM       & $A_{ij} = \sum_k B_{ij}C_{ik}D_{jk}$             & 17 & 62 \\
    \sffamily\bfseries MatTransMul & $y_{i} = \sum_j \alpha A_{ji}^Tx_j + \beta z_i$  & 13 & 50 \\
    \sffamily\bfseries Residual    & $y_{i} = b_i - \sum_j A_{ij}x_j$                 & 9 & 48 \\
    \sffamily\bfseries TTV       & $A_{ij} = \sum_kB_{ijk}c_k$             & 13 & 73 \\
    \sffamily\bfseries TTM       & $A_{ijk} = \sum_lB_{ijl}C_{kl}$         & 11 & 83 \\
    \sffamily\bfseries MTTKRP    & $A_{ij} = \sum_{kl}B_{ikl}C_{jk}D_{jl}$ & 15 & 86\\
    \sffamily\bfseries InnerProd & $\alpha = \sum_{ijk}B_{ijk}C_{ijk}$     & 11 & 115 \\
    \sffamily\bfseries Plus2     & $A_{ijk} = B_{ijk}+C_{ijk}$             & 6 & 163 \\
    \bottomrule
  \end{tabu}
\caption{\label{tab:kernels} The expressions used to evaluate our compiler.}
\label{tab:apps}
\end{table}

\begin{table}
\extrarowheight=1.2pt
  \scriptsize
\begin{tabu}{rrlrS[table-format=3.2e1,round-precision=2,round-mode=places]}
\toprule
\rowfont{\footnotesize\sffamily\bfseries} \multicolumn{2}{c}{App} & Name & Dimensions & {Density} \\
\midrule
\multicolumn{2}{c}{\multirow{3}{*}{\rotatebox{90}{\sffamily\bfseries SpMV}}
\multirow{3}{*}{\rotatebox{90}{\sffamily\bfseries SDDMM}}
\multirow{3}{*}{\rotatebox{90}{\sffamily\bfseries \tiny Mat\ldots Mul}}
\multirow{3}{*}{\rotatebox{90}{\sffamily\bfseries Residual}}} &
bcsstk30~\cite{davis2011university} & $28924\times28924$& 2.4772e-3\\
&& ckt11752\_dc\_1~\cite{davis2011university} & $49702\times49702$&1.34814e-4\\
&& Trefethen\_20000~\cite{davis2011university} & $20000\times20000$&1.3862e-3\\
\midrule
\multicolumn{2}{c}{\multirow{3}{*}{\rotatebox{90}{\sffamily\bfseries Plus3}}} &
random & $800\times800$ & 1.00e-2\\
&&random & $800\times800$& 10.00e-2\\
&&random & $800\times800$& 50.00e-2\\
\midrule
\multirow{4}{*}{\rotatebox{90}{\sffamily\bfseries TTV, TTM}} 
\multirow{4}{*}{\rotatebox{90}{\sffamily\bfseries MTTKRP}} 
&&facebook~\cite{viswanath2009evolution} &$1591\times63891\times63890$&1.13625e-7 
\\\cmidrule{2-5}
& \multirow{3}{*}{\rotatebox{90}{\sffamily\bfseries \tiny InnerProd}}
\multirow{3}{*}{\rotatebox{90}{\sffamily\bfseries Plus2}}
&random & $200\times200\times200$& 1.e-2\\[5pt]
&& random & $200\times200\times200$&10.00e-2\\[5pt] 
&& random & $200\times200\times200$&50.00e-2\\[5pt] 
\bottomrule
\end{tabu}
\caption{The datasets we use for evaluation.}
\label{tab:datasets}
\end{table}

We evaluate \name using the same sparse tensor algebra expressions as in the original tensor compiler (TACO) paper \cite{kjolstad2017tensor}, which can be found in \Cref{tab:kernels}. 
To improve on-chip parallelism, we map Plus3 as an iterated two-input addition because mapping Plus3 natively would only use half of Capstan at a time (performing the first addition and then the second).
CPU baselines are profiled using 128 threads on a four-socket Xeon E7-8890 v3 with a \SI{32}{KiB} L1 data cache, \SI{32}{KiB} L1 instruction cache, \SI{256}{KiB} L2 cache, \SI{46080}{KiB} L3 cache, and \SI{1024}{GiB} RAM. The machine runs Ubuntu 18.04.3 LTS and is clocked at \SI{2494}{MHz}. We compile TACO using GCC 7.4.0 with OpenMP enabled.
GPU baselines are profiled on an AWS EC2 p3.2xlarge	instance with an NVIDIA V100 SXM-2 GPU. The GPU contains \SI{64}{KiB} registers and \SI{12}{KiB} L0 instruction cache per block, \SI{128}{KiB} L1 data cache and shared memory and \SI{2}{KiB} L1 constant cache per  streaming multiprocessor (SM), and \SI{6144}{KiB} L2 cache per GPU. The device RAM is \SI{16160}{MiB}. The GPU has 84 Volta SMs and is clocked at up to \SI{1328}{MHz}. Kernels generated by TACO are compiled with NVCC version 10.0.0. When timing the kernels, we exclude data transfer time between the host and the GPU. Both baselines are run with a cold cache and using a single iteration.

Capstan applications are evaluated with the same cycle-accurate simulator originally used to evaluate Capstan~\cite{rucker2021capstan}. 
The simulator uses an accurate model of the on-chip network, which accounts for delay and throughput constraints~\cite{zhang2019interconnect}; it also uses Ramulator~\cite{kim2016ramulator} to model either four channels of DDR4-2133 or HBM-2E (at 1800 GB/s). The ideal network and memories are more performant since they do not consider latency and any slowdowns, HBM-2E has the highest real-technology memory bandwidth, and DDR4 is slowest. 

\begin{table}
  \scriptsize
  \centering
  \begin{filecontents*}{resources.csv}
App,Par,PCU,pct-pcu,PMU,pct-pmu,MC,pct-mc,Shuf,pct-shuf
SpMV,16,44.0,22.0,41.0,20.5,35.0,43.75,16.0,100.0
Plus3,8,55.0,27.500000000000004,100.0,50.0,58.0,72.5,8.0,50.0
SDDMM,12,163.0,81.5,90.0,45.0,61.0,76.25,0.0,0.0
MatTransMul,16,47.0,23.5,66.0,33.0,36.0,45.0,16,100.0
Residual,16,43.0,21.5,65.0,32.5,36.0,45.0,16,100.0
TTV,16,93.0,46.5,91.0,45.5,67.0,83.75,16.0,100.0
TTM,12,161.0,80.5,89.0,44.5,70.0,87.5,0.0,0.0
MTTKRP,8,140.0,70.0,70.0,35.0,58.0,72.5,0.0,0.0
InnerProd,8,53.0,26.5,155.0,77.5,80.0,100.0,0.0,0.0
Plus2,1,10.0,5.0,23.0,11.5,14.0,17.5,2.0,12.5
\end{filecontents*}
\pgfplotstabletypeset[
    custom pct column/.style={%
        /pgfplots/table/display columns/#1/.style={%
          column type/.add={@{\hskip1ex}}{},
          postproc cell content/.append style={/pgfplots/table/@cell content/.add={\tiny{\normalfont(}}{\,{\normalfont{\tiny\%}})}}},
        },
    col sep=comma,
    numeric type,
    column type={r},
    clear infinite,
    assume math mode,
    fixed,
    fixed zerofill,
    precision=0,
    every row 0 column 8/.style={normpoint},
    every row 0 column 9/.style={normpoint},
    every row 3 column 8/.style={normpoint},
    every row 3 column 9/.style={normpoint},
    every row 4 column 8/.style={normpoint},
    every row 4 column 9/.style={normpoint},
    every row 5 column 8/.style={normpoint},
    every row 5 column 9/.style={normpoint},
    every row 1 column 6/.style={normpoint},
    every row 1 column 7/.style={normpoint},
    every row 2 column 6/.style={normpoint},
    every row 2 column 7/.style={normpoint},
    every row 5 column 6/.style={normpoint},
    every row 5 column 7/.style={normpoint},
    every row 6 column 6/.style={normpoint},
    every row 6 column 7/.style={normpoint},
    every row 7 column 6/.style={normpoint},
    every row 7 column 7/.style={normpoint},
    every row 8 column 6/.style={normpoint},
    every row 8 column 7/.style={normpoint},
    every row 8 column 4/.style={normpoint},
    every row 8 column 5/.style={normpoint},
    every row 2 column 2/.style={normpoint},
    every row 2 column 3/.style={normpoint},
    every row 6 column 2/.style={normpoint},
    every row 6 column 3/.style={normpoint},
    every row 7 column 2/.style={normpoint},
    every row 7 column 3/.style={normpoint},
    display columns/0/.style={string type,
                              column name={},
                              preproc cell content/.code={}},
    assign column name/.style={
      /pgfplots/table/column name={\multicolumn{1}{c}{\sffamily\textbf{#1}}}
    },
    columns/PCU/.style={column name={\#}},
    columns/PMU/.style={column name={\#}},
    columns/MC/.style={column name={\#}},
    columns/Shuf/.style={column name={\#}},
    columns/pct-pcu/.style={column name={\%}},
    columns/pct-pmu/.style={column name={\%}},
    columns/pct-mc/.style={column name={\%}},
    columns/pct-shuf/.style={column name={\%}},
    custom pct column/.list={3,5,7,9},
    every head row/.style={before row={\toprule
    &&\multicolumn{2}{c}{\sffamily\textbf{PCU}}
    &\multicolumn{2}{c}{\sffamily\textbf{PMU}}
    &\multicolumn{2}{c}{\sffamily\textbf{MC}}
    &\multicolumn{2}{c}{\sffamily\textbf{Shuf}}
    \\\cmidrule(lr){3-4}
    \cmidrule(lr){5-6}
    \cmidrule(lr){7-8}
    \cmidrule(lr){9-10}
    }, after row=\midrule},
    every last row/.style={after row=\bottomrule},
    ]{resources.csv}
  \caption{Capstan resources required by our compiled kernels. The specific limiting resource(s) are shown in bold type.}
  \label{tab:resource}
\end{table}

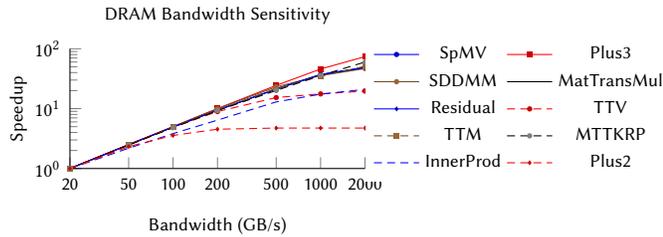
\begin{figure}
  \centering
  \pgfplotstableread[col sep=comma]{bw.csv}\tabbw
  \begin{tikzpicture}
    \begin{axis}[
    height=1.25in,
  axis x line*=bottom,
  axis y line*=left,
width=0.65\linewidth,
        title = {DRAM Bandwidth Sensitivity},
  ymode=log,
  xmode=log,
    legend style={
      legend columns=2,
      anchor=west,
      at={(1.0,0.5)},
      draw=none
    },
    set layers,
    font=\scriptsize\sffamily,
    xmin=20,
    xmax=2000,
    xtick={10,20,50,100,200,500,1000,2000},
    xticklabels={10,20,50,100,200,500,1000,2000},
    ymin=1.0,
    ymax=100,
    ytick={1,10,100},
    mark options={draw opacity=0, scale=0.5},
        y label style={at={(axis description cs:.2,.5)},anchor=south},
  ylabel={Speedup},
  xlabel={Bandwidth (GB/s)},
    ]
            \addplot table[x=key, y=SpMV]  {\tabbw};\addlegendentry{SpMV}
            \addplot table[x=key, y=Plus2CSR]  {\tabbw};\addlegendentry{Plus3}
            \addplot table[x=key, y=SDDMM]  {\tabbw};\addlegendentry{SDDMM}
            \addplot table[x=key, y=MatTransMul]  {\tabbw};\addlegendentry{MatTransMul}
            \addplot table[x=key, y=Residual]  {\tabbw};\addlegendentry{Residual}
            \addplot table[x=key, y=TTV]  {\tabbw};\addlegendentry{TTV}
            \addplot table[x=key, y=TTM]  {\tabbw};\addlegendentry{TTM}
            \addplot table[x=key, y=MTTKRP]  {\tabbw};\addlegendentry{MTTKRP}
            \addplot table[x=key, y=InnerProd]  {\tabbw};\addlegendentry{InnerProd}
            \addplot table[x=key, y=Plus2CSF]  {\tabbw};\addlegendentry{Plus2}
    \end{axis}

  \end{tikzpicture}
  \caption{Impact of memory bandwidth on performance.}
  \label{fig:bw}
\end{figure}

\begin{table*}
  \scriptsize
  \centering
  \resizebox{\linewidth}{!}{
  \begin{filecontents*}{main.csv}
key,Compiled,SpMV,Plus3,SDDMM,MatTransMul,Residual,TTV,TTM,MTTKRP,InnerProd,Plus2,gmean
Capstan (HBM2E) , No,0.6477361153357459,nan,nan,nan,nan,nan,nan,nan,nan,nan,0.6477361153357459
Capstan (Ideal Net \& Mem) , Yes,0.7664357430098683,0.24359952206533603,0.7808701627120009,0.7529674964283335,0.7520726611342982,0.49002788485526366,0.5713001336962407,0.44016575327906793,0.3496182574920894,0.42129089648986834,0.5218725559938878
Capstan (HBM2E) , Yes,1.0,1.0,1.0,1.0,1.0,1.0,1.0,1.0,1.0,1.0,1.0
Capstan (DDR4) , Yes,12.131298693704053,10.067197033307668,8.326635229626831,12.308886421085276,12.063446324283309,4.920316498478387,9.796003750121136,7.759415796349573,3.27713464632134,1.7249738167835245,7.086182554275855
Plasticine (HBM2E) , No,8.719895587565349,nan,nan,nan,nan,nan,nan,nan,nan,nan,8.719895587565349
V100 GPU , Yes,3.1543688754512487,41.88667138565691,18259.502309176525,3.59078330263975,3.5440941075617816,232.8534939449219,284.4693997490397,6.766041230707772,2.757184442057765,381.3762481312448,41.30897436862488
128-Thread CPU , Yes,27.895071391715323,236.39516964932722,220.28248449614298,376.5188021091476,384.08132909211776,335.98634073334955,8.468570832928974,398.7206314026146,178.3353253470917,59.21966437761349,138.0657536974203
\end{filecontents*}
\pgfplotstabletypeset[
    col sep=comma,
    numeric type,
    column type={r},
   clear infinite,
    assume math mode,
    fixed,
    fixed zerofill,
    precision=2,
    empty cells with={\raisebox{0.2ex}{---}},
    display columns/0/.style={string type,
                              column name={Platform (Memory)},
                              preproc cell content/.code={}},
    display columns/1/.style={string type,
                              column type={c},
                              preproc cell content/.code={}},
    display columns/2/.style={
      column type/.add={>{\columncolor[gray]{.8}}}{}
    },
    assign column name/.style={
      /pgfplots/table/column name={\sffamily\textbf{#1}}
    },
    every head row/.style={before row={\toprule&&\multicolumn{5}{c}{\sffamily\textbf{Matrix Kernels}}&\multicolumn{5}{c}{\sffamily\textbf{Tensor Kernels}}\\\cmidrule(lr){3-7}\cmidrule(lr){8-12}}, after row=\midrule},
    every last row/.style={after row=\bottomrule},
    ]{main.csv}}
  \caption{
    Runtimes (geomean across all datasets) normalized to the compiled HBM-2E version of each application for different backends and memories. We compile to Capstan while TACO compiles to CPUs and GPUs. Only SpMV has handwritten kernels.
    \label{tab:perf}
  }
\end{table*}

All evaluations use the datasets shown in \Cref{tab:datasets}. 
For most 2-D kernels, we use the kernels demonstrated in the original Capstan paper~\cite{rucker2021capstan}.
However, Capstan's bit-vector format does not natively support performant co-iteration on highly sparse (less than about 5\%) tensors.
Therefore, we use random datasets with higher densities for these applications (Plus3, InnerProd, and Plus2); these datasets are designed to match Capstan's assumptions about data distribution.
With different backends~\cite{hegde2019extensor,dadu2019towards}, the ideal sparsity ratio would change.

For Plus3, we generate two additional datasets by rotating the input matrix's columns right by one and two.
For Plus2 and InnerProd, we generate one additional dataset by rotating the even coordinates on the last tensor dimension by one. 
Other than the SDDMM and MTTKRP matrices $C$ and $D$ (which are dense), we use CSR (or compressed sparse column (CSC) for MatTransMul) formats for all 2-D matrices. We use compressed sparse fiber (CSF) for most 3-D tensors and a CSR-like uncompressed-compressed-compressed format for InnerProd and Plus2. 

The above formats are used for all platforms except for the GPU baseline which has different result tensor formats. 
Input tensor formats for GPU baseline are the same as above, but output tensors are fully dense. This is because TACO currently does not support sparse output tensors for their GPU backend.

\subsection{Resource Consumption}

For reference, Capstan is built as a grid of 200 vectorized compute units (PCUs) and 200 memory units (PMUs) that are ringed by 80 memory controllers (MCs), as shown in \Cref{fig:capstan}. Each PCU has six stages and 16 vector lanes that can perform fixed or floating-point operations. Each PMU has 16 banks of 4096 32-bit words, supporting one read and one write per bank every cycle. Capstan natively supports sparsity within its architecture by including sparse on-chip memory scheduling, fixed- and floating-point atomics, and sparse iteration over compressed bit-vector data structures. Additionally, 16 shuffle networks enable sparse accesses beyond the 16 lanes of a single PMU, but when used they limit the total number of outer-parallel iterations to 16.

To make good use of Capstan's hardware, a compiler must be able to extract parallelism at both an inner-loop (vectorization) and outer-loop (cross-PCU) level.
Outer-loop parallelism is harder to extract because it requires the compiler to explicitly distribute memories, including FIFOs which must be precisely enqueued and dequeued, across physically unrolled partitions.
Based on the nature of Capstan's distributed compute and memory resources, it is unlikely that an application could use 100\% of all on-chip resources.
Limiting resources vary, but all applications except Plus2 make good use of resources via outer parallelization because they approach a limit in at least one dimension (compute, memory, or shuffle-network ports).
By hitting physical resource limits, the compiler ensures that users are able to take full advantage of Capstan.

One of the key factors in RDA (and GPU) out-performance is a high-bandwidth memory system. Capstan has support for either HBM-2E and DDR4 random access memory.
\Cref{fig:bw} shows that our applications (excepting Plus2, which is not outer-parallelized) are able to make good use of DRAM bandwidth as well.
This is a direct result of the decoupled access-execute memory model: factoring out off-chip memory accesses into large, bulk loads and stores exposes significant memory parallelism.

\begin{figure}
  \centering
  \pgfplotstableread[col sep=comma]{main_tex.csvT}\tabmain
  \begin{tikzpicture}
    \begin{axis}[
  ybar=0pt,
    ymode=log,
    bar width=3.0pt,
    ymin=0.1,
    height=1.25in,
  axis x line*=bottom,
  axis y line*=left,
width=\linewidth,
    xtick=data,
    xticklabels from table={\tabmain}{key},
    yticklabel style={font=\footnotesize},
    xticklabel style={font=\scriptsize,rotate=45,anchor=east},
    legend style={
      legend columns=3,
      anchor=north east,
      at={(1.0,1.0)},
      font={\footnotesize},
      draw=none
    },
                  legend image code/.code={%
                    \draw[#1] (0cm,-0.1cm) rectangle (0.6cm,0.1cm);
                },
    log origin=infty,
      ylabel={\footnotesize Norm. Runtime},
    ]
\addplot table[x expr=\coordindex, meta=key, y=Compiled] {\tabmain};\addlegendentry{Capstan};
\addplot table[x expr=\coordindex, meta=key, y=GPU] {\tabmain};\addlegendentry{GPU};
\addplot table[x expr=\coordindex, meta=key, y=CPU] {\tabmain};\addlegendentry{CPU};
    \end{axis}

  \end{tikzpicture}
  \caption{\label{fig:perf-bar}
  Generated kernel performance across three platforms normalized to Capstan. Our system is used to compile to Capstan and TACO is used to compile to the CPU and GPU. }
\end{figure}
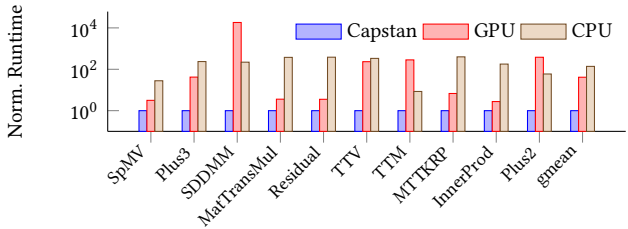

\subsection{Case Study: Sparse Matrix-Vector Multiplication}
\label{sec:spmv}

Sparse matrix-vector multiplication (SpMV) is the simplest of our applications and the only one where a handwritten Spatial implementation exists, as it is significant work to hand-implement a sparse tensor algebra kernel. A comparison of SpMV across all platforms is in the first column of \Cref{tab:perf}. The Capstan and Plasticine rows from \Cref{tab:perf} are handwritten Spatial SpMV kernels from \cite{rucker2021capstan, plasticine} respectively. All Capstan and Plasticine rows in \Cref{tab:perf} that are not compiled (where the Compiled column is No) are hand-written Spatial SpMV kernels. All of our SpMV evaluations use CSR matrices.

SpMV is simpler, making it is easy to parallelize; therefore, SpMV applications compiled by \name have a speedup relative to the CPU and GPU that is lower than other applications. However, the version of SpMV run in the original Capstan paper (Capstan, uncompiled) is more optimized than the compiled version (Capstan, compiled) because the code generated by \name uses the shuffle network (shown by SpMV$\times$Shuf in \Cref{tab:resource}) to coordinate parallel accesses to the input vector and the handwritten Capstan SpMV does not. Instead, the handwritten Capstan SpMV duplicates the input vector, which avoids shuffle-network contention and permits outer-parallelization beyond the shuffle network's limit of 16. We expect that these additional optimizations can be automated in the future, but for now they demonstrate that a dedicated hardware expert can get better performance compared to \name, albeit at the cost of significant development effort.

To demonstrate that \name both increases programmer productivity and decreases development effort in targeting Capstan, we compare the lines of code (LOC) of the handwritten Spatial SpMV kernel against the \name kernel for Capstan. The compiled SpMV kernel uses 10 input LOC total---a 76\% decrease from the 52 lines of Spatial required for the handwritten version. Moreover, we believe that the input code to \name is simpler to write and to port to new architectures. The code required for \name includes: 3 LOC for the tensor formats, 2 LOC for the algorithm, 4 LOC for the scheduling transformations, and 1 LOC to compile and output our kernel. With the use of an auto-scheduler, the number could be cut down from 10 to 6 LOC due to the removal of the user-provided schedule. These numbers support the use of our compiler as a programmer productivity tool for creating arbitrary sparse tensor kernels on accelerators. 

\subsection{Tensor Algebra Expression Performance}

Performance results for all platforms and applications are shown in \Cref{tab:perf}, with a subset of results for only \name compiled Capstan, the TACO compiled GPU, and the TACO compiled CPU shown in \Cref{fig:perf-bar}. The applications compiled by \name are, on average, 138\texttimes\ faster than CPU baselines, and 41\texttimes\ faster than GPU variants. These performance benefits motivate using RDAs---and thus a compiler to target RDAs. 

Our TACO GPU baseline performance is significantly worse than both the literature~\cite{senanayake2020} and compiled Capstan because TACO does not natively support sparse tensor outputs. Most of the time is spent zero initializing the fully dense result tensor---which is often extremely large---in device memory. Because Capstan is designed to outperform the GPU for sparse applications, it may seem counter-intuitive that the GPU speedup for MTTKRP is relatively low. However, these kernels have a dense dimension that the GPU can vectorize.

A comparison of compiled Spatial applications to handwritten Spatial implementations for Capstan across the applications is not possible, since handwritten kernels do not exist, except for SpMV. They do not exist because each application must be written by a Spatial, Capstan, and application/sparsity domain expert. Additionally, a handwritten implementation is slower to write since the Spatial programming abstraction is at a lower-level. In order to hand-write correct Spatial for Capstan, the domain expert must have deep knowledge of the Spatial-to-Capstan compilation process through both the Spatial and SARA compilers. The fact that our system is able to compile to applications where a hand-written Spatial implementation does not exist motivates the use of our automated compiler to generate the long tail of sparse tensor algebra kernels, which have practical applications, including deep neural networks (SDDMM), scientific computing (Plus3), linear least squares (Residual), and alternating least squares (MTTKRP).

\section{Related Work}
\label{sec:related-work}

\name is, to the best of our knowledge, the first compiler to compile a high-level sparse tensor algebra expression language to a general sparse domain-specific accelerator. There is, however, prior work on sparse tensor algebra systems targeting CPUs and GPUs, compilers for dense image processing and neural network applications that target dense domain-specific architectures, other sparse domain-specific hardware that provides alternative targets for a compiler like ours, and different methodologies for programming these sparse DSAs.

\paragraph{Sparse Tensor Algebra Systems and Compilers} 
Several compilers have been proposed for sparse tensor algebra, but 
these compilers target CPUs~\cite{bik1993,kotlyar1997a,venkat2015chillie,kjolstad2017tensor,henry2021} or GPUs~\cite{senanayake2020}, whereas our system compiles to domain-specific sparse data\-flow hardware. We chose to extend TACO with a compilation path to target RDAs since it contains input languages and IRs that support general tensor compression formats, optimizations, and expressions. TACO, unlike our system, defines co-iteration as only the intersection of tensor coordinates. In order to support all tensor algebra expressions, TACO uses an iteration lattice IR to decompose all unions of coordinates into disjoint intersections and then emits code that performs a multi-way merge strategy, 
whereas \name emits scanners through logical operations on bit vectors.

\paragraph{Sparse Domain-Specific Hardware}

Other architectures have been proposed for single-function sparse acceleration, including matrix-vector multiplication~\cite{han2016eie} and matrix-matrix multiplication~\cite{srivastava2020matraptor,zhang2021gamma,zhang2020sparch,pal2018outerspace}. 
Prior work has also mapped algorithms like SpMV~\cite{shan2010fpga,umuroglu2014energy,zhang2009fpga,grigoras2015accelerating} to FPGAs. 
Because the compiler would not reconfigure the FPGA itself, pre-defined FPGA implementations would be exposed to the programmer in the same manner as an ASIC.
From a compiler perspective, these algorithms can all be considered as special-purpose lowering cases.
For example, a sparse matrix-matrix multiply could be mapped to our \code{accelerate} command.

\paragraph{Mapping Sparsity to DSAs}
We focused our evaluation on Capstan~\cite{rucker2021capstan} because it is a flexible RDA with an easy-to-understand programming model: it supports vectorized sparse iteration with composable parallel patterns.
However, sparse iteration spaces are a general representation, and \name could target any reconfigurable sparse accelerator with appropriate backend modifications.
The SPU~\cite{dadu2019towards} and ExTensor~\cite{hegde2019extensor} are two recent sparse DSAs with different programming models than Capstan.
Both have tiled architectures with explicit on-chip memory accesses, but they have different methods for combining sparse data.

The SPU uses a stream-join programming abstraction to combine sparse indices and a custom RDA fabric to perform the intersection operations.
The core of stream-join is a conditional-dequeue abstraction with FIFO inputs: to lower to this abstraction, our compiler would have to rewrite CIN traversal using FIFO dequeues (instead of the pointer bumps and memory accesses used for CPU backends).
Similarly, ExTensor uses a programming model based on hierarchical tensor intersectors; these are designed to provide high throughput for direct iteration on sparse data structures.
However, any compiler would still need to handle low-level mapping, including coordinating memory transfers.

\section{Conclusion}
\label{sec:conclusion}

We have described the first compiler that automatically compiles arbitrary sparse tensor algebra kernels to sparse reconfigurable dataflow accelerator. Since architects are developing many accelerators for sparse computations, we expect \name to be the first of many to target sparse accelerators. We expect its design---leveraging the separation of application code from both data representation \textit{and} schedules to independently map data to memories and (sparse) operations to sparse data combiners and compute units---will influence future domain-specific compiler designs. Finally, we plan to further extend our compiler path into a compiler toolkit that can be used to create sparse tensor algebra compiler backends for a host of emerging domain-specific sparse hardware.

\section{Acknowledgements}

We thank Willow Ahrens, Stephen Chou, Scott Kovach, Aviral Pandey, and Rohan Yadav for their helpful feedback and discussion. Olivia Hsu was supported by an NSF GRFP Fellowship, and Alexander Rucker was supported by a Stanford Graduate Fellowship. This work was supported in part by the NSF under grant numbers 1937301, 2028602, CCF-1563078, and 1563113.  
This research was also supported in part by the Google Research Scholar program, Facebook Systems for ML research award, and Stanford Data Analytics for What's Next (DAWN) Affiliate Program. Any opinions,
findings, and conclusions or recommendations expressed in this material are those of the authors
and do not necessarily reflect the views of the aforementioned funding agencies.

\clearpage
\bibliography{references}

\end{document}